\documentclass[acmlarge]{acmart}
\makeatletter
\newcommand{\confshort}{\acmConference@shortname}
\newcommand{\conffull}{\acmConference@name}
\newcommand{\confdate}{\acmConference@date}
\newcommand{\confloc}{\acmConference@venue}
\AtBeginDocument{
  \fancypagestyle{firstpagestyle}{
    \fancyhead{}%
    \fancyfoot[C]{}%
  }
  \fancyhf{}
  \fancyhead[LO]{\@headfootfont\shorttitle}%
  \fancyhead[RE]{\@headfootfont\@shortauthors}%
  \fancyhead[LE]{\@headfootfont\footnotesize \confshort, \confdate, \confloc}%
  \fancyhead[RO]{\@headfootfont\footnotesize \confshort, \confdate, \confloc}%
  \fancyfoot[C]{}%
}
\makeatother
\acmBooktitle{\conffull\@ (\confshort), \confdate, \confloc}

\AtBeginDocument{%
  }
\copyrightyear{2026}
\acmYear{2026}
\setcopyright{cc}
\setcctype{by}
\acmConference[FAccT '26]{The 2026 ACM Conference on Fairness, Accountability, and Transparency}{June 25--28, 2026}{Montreal, QC, Canada}
\acmBooktitle{The 2026 ACM Conference on Fairness, Accountability, and Transparency (FAccT '26), June 25--28, 2026, Montreal, QC, Canada}
\acmDOI{10.1145/3805689.3812406}
\acmISBN{979-8-4007-2596-8/2026/06}
\usepackage{booktabs}
\usepackage[dvipsnames]{xcolor}
\usepackage{makecell}
\usepackage{tabularx}
\usepackage{caption}
\usepackage{subcaption}
\usepackage{multirow}
\usepackage{float}
\usepackage{algorithm}
\usepackage{algpseudocode}

\usepackage{amssymb}
\usepackage{amsmath}
\usepackage{placeins}
\usepackage[table, dvipsnames]{xcolor}
\begin{document}

%%
%% The "title" command has an optional parameter,
%% allowing the author to define a "short title" to be used in page headers.
\title[Democratizing News Recommenders]{Democratizing News Recommenders: Modeling Multiple Perspectives for News Candidate Generation with VQ-VAE}

%%
%% The "author" command and its associated commands are used to define
%% the authors and their affiliations.
%% Of note is the shared affiliation of the first two authors, and the
%% "authornote" and "authornotemark" commands
%% used to denote shared contribution to the research.
\author{Hardy}
\email{hardy@ims.uni-stuttgart.de}
\orcid{0000-0003-0629-0339}
\affiliation{%
  \institution{University of Stuttgart}
  \city{Stuttgart}
  \state{Baden-Württemberg}
  \country{Germany}
}
\affiliation{%
  \institution{Universitas Mikroskil}
  \city{Medan}
  \state{North Sumatra}
  \country{Indonesia}
}

\author{Sebastian Pad{\'o}}
\email{pado@ims.uni-stuttgart.de}
\orcid{0000-0002-7529-6825}
\affiliation{%
  \institution{University of Stuttgart}
  \city{Stuttgart}
  \state{Baden-Württemberg }
  \country{Germany}
}

\author{Amelie W{\"u}hrl}
\email{amwy@itu.dk}
\orcid{0009-0008-3382-0423}
\affiliation{%
  \institution{IT University of Copenhagen}
  \city{Copenhagen}
  \country{Denmark}}

\author{Tanise Ceron}
\email{tanise.ceron@unibocconi.it}
\orcid{0009-0002-4845-2789}
\affiliation{%
  \institution{Bocconi University}
  \city{Milan}
  \country{Italy}
}

%%
%% By default, the full list of authors will be used in the page
%% headers. Often, this list is too long, and will overlap
%% other information printed in the page headers. This command allows
%% the author to define a more concise list
%% of authors' names for this purpose.

%%
%% The abstract is a short summary of the work to be presented in the
%% article.
\begin{abstract}

News Recommender Systems (NRS) play a central role in society. They shape what users read, whose perspectives they encounter, and influence public discourse. Yet their design is value-laden: intentionally or not, NRS can embed undesired values in their recommendation procedures, such as exclusion of underrepresented voices or favoring specific viewpoints, which conflict with democratic goals. Despite this, existing solutions lack an intervention mechanism that controls these values. Therefore, we introduce an approach that parameterizes NRS with the goal of intervening in such values and promoting different democratic goals in models. 

We propose Aspect-Aware Candidate Generation (\texttt{A2CG}), a normatively configurable procedure for the candidate generation stage of NRS which allows designers to explicitly shape diversity in the recommendations. \texttt{A2CG} introduces diversity at the start of the recommendation system rather than only re-ranking candidates as a post-processing step. \texttt{A2CG} diversifies recommendations by representing articles along different diversity aspects: sentiment, political leaning, topic, and media framing. User interests are encoded over these aspects using a Vector Quantized VAE, and a decoder-only model predicts the next article aspects the users are likely to engage with. To broaden exposure to perspectives, \texttt{A2CG} injects diversity at retrieval by selectively flipping aspects in the predicted query, allowing candidate diversity to be tuned for specific democratic models. 

Our method enables qualitatively different normative configurations that existing NRS cannot express, and unlike baselines with fixed structural biases, \texttt{A2CG} allows continuous calibration between democratic ideals without retraining. Our findings empirically show that \texttt{A2CG} generates novel, diverse, and serendipitous candidates while providing explicit, parameter-driven control over the trade-off between personalization and democratic alignment. Rather than aiming for pointwise superiority over existing methods, \texttt{A2CG}’s key contribution lies in its controllability and ability to express flexible normative configurations.
\end{abstract}




% \received{20 February 2007}
% \received[revised]{12 March 2009}
% \received[accepted]{5 June 2009}

\maketitle

\section{Introduction}

\textit{News Recommender Systems} (NRS) are widely implemented in online news outlets to offer personalized news selections to users \citep{thurmanFuturePersonalizationNews2012, bastianExplanationsNewsPersonalisation2020}, proposing to reduce information overload \citep{resnickGroupLensOpenArchitecture1994}. Such systems are primarily designed for optimizing click-through rates and user engagement \citep{liContextualBanditApproachPersonalized2010} by predicting news articles that are interesting to the users \citep{thurmanFuturePersonalizationNews2012, bastianExplanationsNewsPersonalisation2020}. This objective has the potential to lead to undesirable outcomes \citep{golbeck2020optimizing}, such as echo chambers \citep{nguyenEchoChambersEpistemic2020}, selective exposure \citep{frey1986recent,hart2009feeling}, or popularity bias \citep{zhang2021causal}. By optimizing for narrow engagement metrics,  their predictions often do not foster democratic values \citep{helbergerDemocraticRoleNews2019}, such as informed public deliberation and active citizenship.

A number of NRS approaches aim at addressing these problems, focusing on designing diversity-aware or recommendation algorithms that maximize topical variety or novelty within user sessions \citep{gharahighehiDiversificationSessionbasedNews2023, ianaTrainOnceUse2024, qiHieRecHierarchicalUser2021}. In theory, however, diversity has many definitions, as it can refer to diversity of sources, narrators, or perspectives in terms of viewpoints \citep{loecherbach2020unified}. These properties admit several definitions and metrics, often leading to competing optimization goals \citep{wu_result_2024} because of conflicting interests, for example, showing a wider variety of topics instead of different viewpoints on the same topic. Therefore, designing a democratically aligned NRS involves two interconnected goals: (1) introducing diversity, and (2) doing so in a way that satisfies the needs from different citizen archetypes \citep{helbergerDemocraticRoleNews2019,vrijenhoek2021recommenders}. In practice, NRS should introduce new or underrepresented perspectives while accounting for different archetypes in a democratic society. Following \citet{helbergerDemocraticRoleNews2019}, this means calibrating diversification across four main citizen models -- \textit{Liberal}, \textit{Participatory}, \textit{Deliberative}, and \textit{Critical} -- for instance, the \textit{Liberal} model favors closer alignment with users’ views, whereas the \textit{Participatory} model emphasizes content that challenges them.

Concerning point (1), NRS often rely on a simplified proxy of perspectives to promote diversity, such as topic variety or keyword differences \citep{wangNewstopicRecommenderSystem2018, leeNewsRecommendationTopicEnriched2020}. Such proxy-based approaches fall short of operationalization without explicit grounding in normative theory. As such, existing work remains largely at conceptual or evaluation level, with no concrete operationalization of NRS that support different democratic models using more sophisticated proxies for perspectives. Regarding point (2), many systems improve diversity by re‑ranking a pre‑retrieved candidate list \citep{wangNewstopicRecommenderSystem2018, qiHieRecHierarchicalUser2021, gharahighehiDiversificationSessionbasedNews2023, ianaTrainOnceUse2024}, rather than introducing new candidates beyond what has already been retrieved. This post-processing approach inherently limits diversification. As \citet{dufraisse2022don} demonstrate, re-ranking alone fails to meaningfully increase diversity when the candidate pool is homogeneous. This limitation highlights a fundamental challenge: implementing democratic principles in NRS should start with a candidate pool that covers a broad spectrum of viewpoints and that can meet the criteria for different democratic models more easily. 

\begin{figure*}[t!]
  \Description{Overview of the A2CG framework.}
  \centering
  \includegraphics[width=0.9\linewidth]{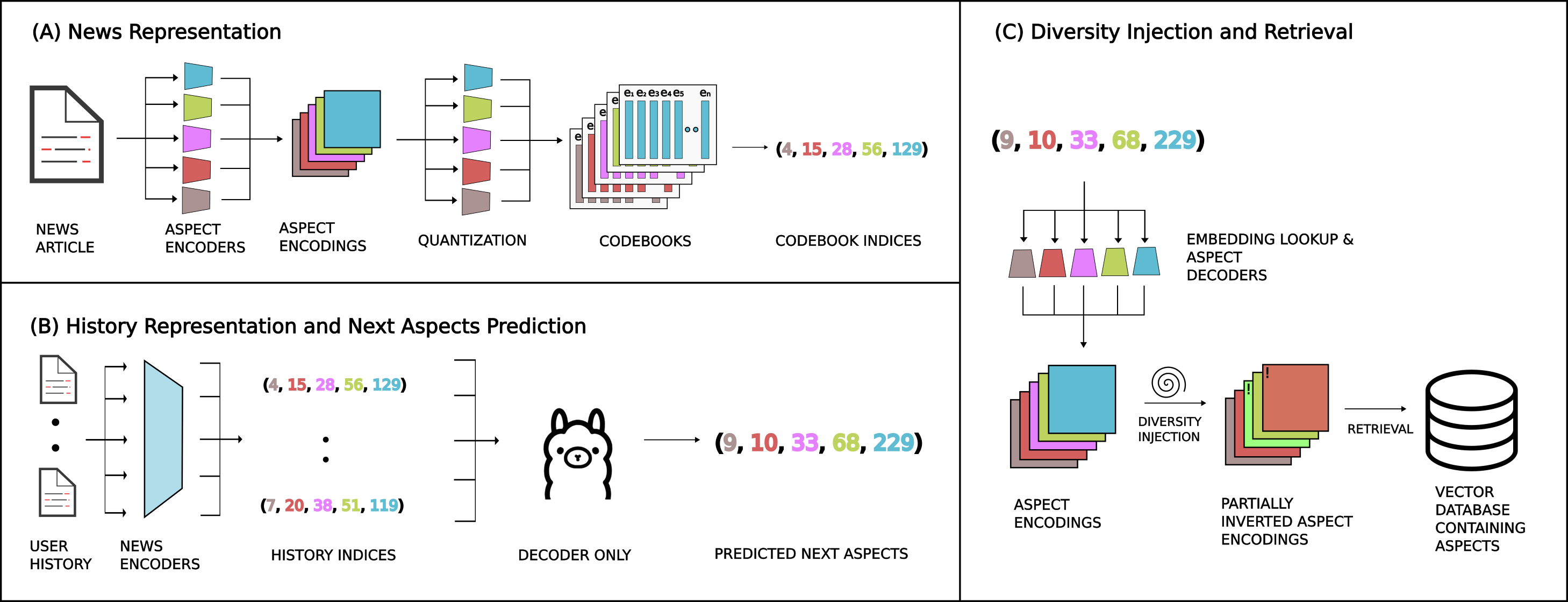}
  \caption{Overview of the \texttt{A2CG} framework. 
  (A) News articles are encoded and quantized into discrete indices. 
  (B) Quantized user history is used to predict the next articles. 
  (C) Predicted articles are diversified before candidate retrieval.}
  \label{fig:overview}
\end{figure*}

In this paper, we address both points (1) and (2) with a framework called Aspect-Aware Candidate Generation (\texttt{A2CG})\footnote{Code is available in \url{https://github.com/blodstone/DemocraticRecSys}}. The framework, shown in Figure \ref{fig:overview}, promotes distinct democratic models at the \textit{candidate generation stage} through an intervention mechanism. By introducing diversity constraints at this early stage, we can ensure that a broader and more representative set of news articles is considered for recommendation. Furthermore, we derive user archetypes from \texttt{A2CG} by varying its parameter sets. In our work, archetypes are user reading preferences that can be mapped onto the requirements for the candidate generations in the different democratic models, allowing us to conveniently calibrate our model depending on the user's fitting democratic model. To the best of our knowledge, this is the first approach in the candidate generation stage that focuses on democratic aims.

To realize this framework in practice, \texttt{A2CG} represents article content as a bundle of distinct, manageable ``aspects'' (e.g., political leaning, news category, sentiment) which are proxies for identifying perspectives in texts. We use these aspects to steer the candidate generation process. We utilize Vector Quantized Variational Autoencoders (VQ-VAE) \citep{oordNeuralDiscreteRepresentation2018} to create aspect-specific discrete latent spaces, referred to as \emph{codebooks}, which store a finite set of learnable embedding vectors used to represent each aspect. This approach enables a decoder-only model \citep{touvronLLaMAOpenEfficient2023a} to learn user preferences over a discrete, multi-faceted representation of articles and generate novel candidates by predicting a new sequence of unseen aspects. This architecture enriches the article representations for an efficient retrieval process while still effectively capturing user history. By mapping these rich, quantized representations to a vector space and using a fast nearest-neighbor search algorithm, we can retrieve a diverse set of candidates. While nearest-neighbor search is a standard technique, \texttt{A2CG} allows us to go beyond simple item similarity and retrieve a broader, more diverse set of articles in an efficient offline framework. 

We comprehensively evaluate our approach along the following research questions:
\begin{enumerate}
    \item \textbf{To what extent does \texttt{A2CG} enable parameterized intervention over democratic goals?} 
    We benchmark vanilla \texttt{A2CG} under different normative parameter settings against a range of baseline methods, including those optimized for traditional and normative diversity metrics. Our analysis examines how changes in \texttt{A2CG}'s intervention parameter systematically alter recommendation outcomes, demonstrating its ability to modulate exposure in line with different democratic objectives -- capabilities that baseline systems lack.
    \item \textbf{How do different democratic configurations of \texttt{A2CG} reshape exposure for distinct user archetypes under the four democratic models \citep{karppinenUsesDemocraticTheory2012, helbergerDemocraticRoleNews2019,vrijenhoekRADioIntroductionMeasuring2024}}
    We analyze how \texttt{A2CG} affects recommendation outcomes for different user-content archetypes across four democratic models. This evaluation examines how \texttt{A2CG}'s tunable intervention mechanism enables model-specific adjustments in perspective exposure. We highlight its ability to increase exposure to diverse content while promoting distinct democratic goals.
    \item \textbf{How does \texttt{A2CG}'s multi-aspect approach balance the trade-off between diversity and personalization?}
    We investigate how \texttt{A2CG}'s diversity injection mechanism influences the trade-off between personalization and diversity. Our analysis examines the trends and specific metrics that are affected by this process.
\end{enumerate}

Our key contributions are: 
\begin{enumerate}
    \item A novel multi-aspect framework for diverse candidate generation at the underexplored retrieval stage of recommendation systems. Unlike re-ranking approaches constrained by a fixed candidate pool, our approach injects diversity into the multi-aspect representation itself, enabling continuous and parameter-driven calibration between personalization and democratic alignment without retraining. 
    \item A generative approach with a VQ-VAE and a decoder-only model to create novel and unseen bundles of aspects. These generated bundles serve as a unique representation of the next article a user is likely to read. 
    \item The operationalization of four models of democracy as a theoretical framework for our model analysis.
    \item The development of a new set of strong baselines -- each exhibiting fixed structural biases toward specific democratic ideals -- alongside a comprehensive archetype analysis demonstrating A2CG's flexibility to target any democratic model through parameter adjustment alone.
\end{enumerate}

\section{Related Work}
\paragraph{News Recommendation System (NRS)} 
Recent work in NRS uses neural content-based models to  personalize news recommendations. These systems typically employ a two-tower architecture, consisting of a  news encoder and a user history encoder \citep{wuPersonalizedNewsRecommendation2023,ianaTrainOnceUse2024,huangComprehensiveSurveyRetrieval2025}. The news encoder extracts semantic representations from textual attributes such as titles, categories, and bodies, often using Transformers \citep{vaswaniAttentionAllYou2017a}. Simultaneously, the user encoder aggregates these representations from a user's clicked news history to capture their interests. MANNeR \citep{ianaTrainOnceUse2024} employs a modular  architecture: A content module captures the user history, while an aspect module captures a single aspect of the article (e.g., sentiment or politics). The two modules are independently trained but later used together during inference. \texttt{A2CG} uses the principle behind aspect modules for aspect prediction while applying the two-tower architecture for retrieval.

\paragraph{Retrieval-focused RS} Large-scale recommendation systems (RS) use multi-stage pipelines that include a retrieval (or candidate generation) stage followed by ranking and re-ranking stages \citep{huangComprehensiveSurveyRetrieval2025} to narrow millions of items into relevant subsets, though research is largely focused on non-news domains like e-commerce\citep{zhouContrastiveLearningDebiased2021}, books \citep{shiSPARCSoftProbabilistic2025}, videos\citep{liuRecFlowIndustrialFull2024}, games store\citep{zhouContrastiveLearningDebiased2021}, or movies \citep{qinRankFlowJointOptimization2022}. In the news domain, \citet{lianBetterRepresentationLearning} address retrieval with a Deep Fusion Model combining multi-level feature interactions with attentive signal weighting for low-latency matching. \citet{abbarRealtimeRecommendationDiverse2013} uses a two-stage pipeline which uses Jaccard similarity and subsequent diversity maximization. The system employs the Greedy Max-Min algorithm to select a subset of news articles that maximizes diversity distance, achieving a 2-approximation of the optimal set. However, since neither \citet{abbarRealtimeRecommendationDiverse2013} nor \citet{lianBetterRepresentationLearning} published their datasets, and most existing public datasets \citep{gullaAdressaDatasetNews2017, wuMINDLargescaleDataset2020a, kruseEBNeRDLargeScaleDataset2024a} are designed for ranking with pre-filtered candidate items, a gap remains. To bridge this gap, \texttt{A2CG} adapts \texttt{MIND} into a retrieval benchmark by treating all articles published daily as a global candidate pool.

\paragraph{Diversity Aware RS} Much of the early work on diversity in the retrieval stage of RS focuses on descriptive diversity, which emphasizes the variety of items or categories. \citet{abbarRealtimeRecommendationDiverse2013} formalize retrieval as a balance between relevance and intra-list diversity, ensuring variety during the initial phase. \citet{zhouContrastiveLearningDebiased2021} focus on aggregate diversity via contrastive learning to counter popularity bias across the entire user base. While SPARC \citep{shiSPARCSoftProbabilistic2025} incorporates diversity to explore long-tail interests, they treat it as a technical objective rather than a normative democratic goal. Work in the retrieval stage of the news domain, such as \citet{lianBetterRepresentationLearning}, focuses only on relevance. In the ranking stage, \citet{ianaTrainOnceUse2024} uses a descriptive type of diversity, specifically aspect-based diversity, which is defined as the normalized entropy of the distribution of an aspect in the recommendation list. However, the academic discourse has shifted toward a more nuanced understanding known as normative diversity. This concept is rooted in democratic theory and the societal role of journalism, advocating for user exposure to a representative range of viewpoints and perspectives rather than just topics \citep{helbergerDemocraticRoleNews2019, karppinenUsesDemocraticTheory2012}. The RADio framework \citep{vrijenhoekRADioIntroductionMeasuring2024}  proposes to evaluate normative diversity, bridging the gap between theory and  technology. \texttt{A2CG} extends the \texttt{RADio} framework to support different democratic models, introducing a novel approach to measuring diversity in the retrieval stage of recommendation systems.

\section{Methods}

\subsection{Candidate Generation}
We consider the candidate generation problem with a large set of possible news article candidates, $X = \{x_i\}_{i=1}^{|X|}$, as the starting point for our formulation. Each user $u$ has a history of previously read articles, $H_u \subset X$. The  objective is to retrieve candidate articles, $C \subset X$ with $|C| \ll |X|$ such that $C$ is relevant to the user's history while also being diverse.

Our proposed method, \texttt{A2CG}, centers on learning a representation of what the user is likely to read next, $q_u$, which serves as a query for retrieving $C$. However, due to the inherently discrete nature of news items and the vast space of possible candidates, it is not possible to directly predict the target article. To address this, we instead focus on an approach that generates the \textit{abstract concept} of the article by combining different aspects. 

\subsubsection{Multi-Aspect Encoding of An Article}
\label{subsec:multi-aspects-embedding}

The multi-aspect encoding of an article is depicted in the first half of Figure \ref{fig:overview}(A). We represent the articles using five different aspects: article content, sentiment, category, political leaning, and frames. However, our framework can be applied to any set of aspects. 

Formally, an \emph{aspect}, denoted by $\mathcal{A}$, represents a specific characteristic or dimension of an article’s content that can be used to differentiate its viewpoint. Let $\mathcal{A} = { \mathcal{A}_1, \mathcal{A}_2, \dots, \mathcal{A}_K }$ denote the set of all aspects considered, where each $\mathcal{A}_k$ corresponds to a particular attribute $k$ (e.g., sentiment, political leaning, or framing). In our dataset, each news article $x_i \in \mathcal{D}$ is annotated with labels for four of the five aspects, and our goal is to learn an aspect-specific embedding function for each aspect. For the \textbf{article content} aspect, the encoding function operates directly on the raw text of the article’s title and abstract, without relying on an explicit label. We use an adapted Sentence Transformer model \citep{reimersSentenceBERTSentenceEmbeddings2019} with ModernBERT \citep{warnerSmarterBetterFaster2024} as the base encoder to embed the \textbf{article content} aspect.
For the remaining four aspects, we fine-tune each aspect-specific embedding 
with supervised contrastive learning \citep{khoslaSupervisedContrastiveLearning2021, ianaTrainOnceUse2024}. We use the  article content aspect, $k$, as base embedding:

\begin{equation}
\mathcal{L}^{(k)} 
= \sum_{i \in I} \mathcal{L}_{i}^{(k)} 
= - \sum_{i \in I} \frac{1}{|P^{+}(i)|} \sum_{p \in P^{+}(i)} 
\log \frac{\exp \big( z_i^{(k)} \cdot z_p^{(k)} / \tau \big)}
{\sum\limits_{a \in A(i)} \exp \big( z_i^{(k)} \cdot z_a^{(k)}) / \tau \big)},
\end{equation}
where $z_i^{(k)} = f_k(x_i)$ denotes the embedding of an article $x_i$ under aspect-specific encoder $f_k$, $\tau > 0$ is a temperature parameter, $i \in I$ is the index set of our training dataset, $P^{+}(i)$ is the set of \emph{positive} samples sharing the same label $y_k$, and $A(i)$ is the set of comparison samples (positives and negatives).

\subsubsection{Article Quantization}

To train our decoder, we apply quantization, which, as shown in the second half of Figure \ref{fig:overview}a, enables discrete learning. For each aspect $k \in \{1, \dots, K\}$, we train a separate Vector Quantized Variational Autoencoder (VQ-VAE)\citep{oordNeuralDiscreteRepresentation2018} to discretize its continuous space. The VQ-VAE uses vector quantization to map a continuous, high-dimensional input into a discrete set of learned codebook vectors, preserving the aspect information. 

Given the aspect-specific embedding, $z_k \in \mathbb{R}^d$, of aspect $k$, the encoder $E_k: \mathbb{R}^d \rightarrow \mathbb{R}^{d_e}$ outputs a single latent vector, $h$, which is quantized to the nearest entry in the aspect-specific codebook $\mathcal{C}_k = \{ e_j^{(k)} \}_{j=1}^{|\mathcal{C}_k|}$:
$
i_k = \arg\min_{j \in \{1, \dots, |\mathcal{C}_k|\}}  \| h - e_j^{(k)} \|_2
$,
where $i_k$ is the chosen codebook index for aspect $k$ and subsequently $\phi_k(x) = e_{i_k}^{(k)}$ is the quantized representation for aspect $k$.
The VQ-VAE is trained to reconstruct the original aspect embedding $z_k$ from its quantized form while also learning an effective codebook. The training loss has three components:
\allowdisplaybreaks
\begin{align}
\mathcal{L}_\mathrm{VQ}^{(k)} 
= & \underbrace{\| z_k - D_k(\phi_k(x)) \|_2}_{\text{Reconstruction loss: ensure decoded output matches $z_k$}} \\
& + \underbrace{\| \mathrm{sg}[E_k(z_k)] - \phi_k(x) \|_2}_{\text{Codebook loss: updates codebook to match encoder outputs}} \\ &+ \underbrace{\beta \| E_k(z_k) - \mathrm{sg}[\phi_k(x)] \|_2}_{\text{Commitment loss{ encourage encoder outputs to stay near chosen codes}}},
\end{align}
where $D_k$ is the decoder for aspect $k$, $\mathrm{sg}[\cdot]$ is the stop-gradient operator, and $\beta$ is the commitment weight.

The final representation of an article $x$ is the  sequence of the $K$ aspect-wise quantized codebook indices,
$\Phi(x) = \big[i_1, i_2, \dots, i_K \big]$.

\subsubsection{Decoder-Only Modeling over Quantized Aspect Codes}
The final step of our candidate generation (depicted in Figure \ref{fig:overview}(B)) involves modeling user history over the quantized multi-aspect representation of news articles. We use a decoder-only model to learn the conditional probability of the next quantized article representation, given a user's reading history, but any autoregressive LM (including all current  generative Transformer models) can be used. 

Given a user $u$ with a history of articles, $H_u = (\Phi(x_1), \cdots, \Phi(x_k))$, the model is trained to predict the next article's representation, $\Phi(n_{k+1})$. To do this, we first unroll the history into a single, flattened sequence of indices:
\begin{equation}
\begin{split}
S = [\Phi(x_1), \Phi(x_2), \dots, \Phi(x_k)] =
 [i_{1,1}, \dots, i_{1,K}, i_{2,1}, \dots, i_{2,K}, \dots, i_{k,1}, \dots, i_{k,K}]
\end{split}
\end{equation}
where $S$ is the flattened sequence of indices from the user's history.

The model is then trained to maximize the conditional probability of the next token, using a cross-entropy loss:
\begin{equation}
\mathcal{L}_{\mathrm{hist}} 
= - \sum_{t=1}^{|S|-1} 
\log P \big( S_{t+1} \,\mid\, [\mathrm{BoS},  S_{\leq t}]).
\end{equation}

During inference, we condition on the entire history to predict the next $K$ aspect indices.
\begin{equation}
P(\Phi(x_{k+1}) | S) = \prod_{j=1}^{K} P(\Phi(x_{k+1})_j | S, \Phi(x_{k+1})_{<j})
\end{equation}

\subsubsection{Retrieval via Quantized Codebook Embeddings}
Once the decoder model has been trained to predict quantized codes, we perform candidate retrieval (Figure \ref{fig:overview}c) by mapping articles into a vector database using  codebook embeddings. 

\paragraph{Storage and Query Formulation.}
For each article $x_i$ in the set $X$, we embed it $K$ times using the fine-tuned aspect models to obtain $(z_1, z_2, \dots, z_K)$. We then apply Singular Value Decomposition (SVD) to reduce the dimensionality of each embedding. These reduced vectors are then concatenated and stored in a vector database.

At query time, we use the decoder-only model to predict the next $K$ aspect indices. We then retrieve the corresponding codebook embeddings from the 
quantizer, followed by decoding each embedding into its aspect representation:
\begin{equation}
q(H_u) = f(\Phi(x_{k+1})) = \big(e^{(1)}_{j_1}, e^{(2)}_{j_2}, \dots, e^{(K)}_{j_K}\big),
\end{equation}
where $f(\cdot)$ maps each predicted index to its codebook embedding $e^{(m)}_{j_m}$ for aspect $m$ and then decodes it into the final aspect vector.
The predicted aspect index denotes the user-content archetype by mapping the user's historical interaction patterns into a discrete, multi-faceted representation of perspective preferences. For each archetype, we characterize its normative properties by calculating the label proportions of all articles associated with its index.

\begin{algorithm}[tb!]
\caption{Query Formulation with Diversity Injection}
\label{alg:diversity_injection}
\begin{algorithmic}[1]
\Require Predicted aspect indices $\Phi(n_{k+1}) = (j_1, \dots, j_K)$, 
         flip probability $r$, number of aspects to flip $n$
\Ensure Concatenated candidate vector $\mathbf{q}$
\State $\textit{switches} \gets \varnothing$
\If{$\textsc{Random}() < r$}
    \State $\textit{switches} \gets$ \textsc{RandomSample}($\{1,\dots,K\}, n$)
\EndIf
\For{$m \gets 1$ to $K$}
    \State $\mathbf{a} \gets$ \textsc{SVDTransform}(\textsc{f}($j_m$))
    \If{$m \in \textit{switches}$}
        \State $\mathbf{a} \gets -\mathbf{a}$ \Comment{Invert vector direction}
    \EndIf
    \State Append $\mathbf{a}$ to $\textit{concat\_vectors}$
\EndFor
\State $\mathbf{q} \gets$ \textsc{Concatenate}($\textit{concat\_vectors}$)
\State \Return $\mathbf{q}$
\end{algorithmic}
\end{algorithm}

\paragraph{Diversity Injection}
To promote diversity in candidate representations, we introduce a \emph{diversity injection} step (Algorithm \ref{alg:diversity_injection}). Given the predicted aspect indices $\Phi(n_{k+1}) = (j_1, \dots, j_K)$, we decode and reduce each aspect vector and, with probability $r$ (line 2), randomly select $n$ aspects (line 3) to have their vectors inverted (multiplied by $-1$, line 8) before concatenation. This inversion is a deliberate geometric choice: multiplying an aspect embedding by $-1$ preserves its magnitude while systematically redirecting the query away from the predicted aspect during retrieval. Unlike adding noise or resampling from the codebook, inversion preserves the scale and structural relationships between aspects, ensuring that the diversified query remains within the same embedding space. The two parameters $n$ and $r$ provide orthogonal control, where $n$ determines how many aspects are shifted and $r$ controls the probability of any shift occurring. Together they produce the smooth and predictable transitions observed in Figure \ref{fig:four_models}, where increasing either parameter monotonically steers candidates toward broader democratic configurations.
\paragraph{Nearest Neighbor Retrieval. }
Using $q$ as a query vector, we perform a nearest neighbor search over the database of concatenated article embeddings. Specifically, we employ the Hierarchical Navigable Small-World (HNSW) algorithm \citep{malkovEfficientRobustApproximate2020} to retrieve the top-$k$ most similar article vectors,
$C = \mathrm{HNSW}\big( q, \{ E(x_i) \}_{i=1}^{|N|}, k \big)$,
where $C \subset N$ contains the $k$ retrieved candidates. HNSW is a fast graph-based approximate nearest neighbor whose runtime scales logarithmically. These retrieved articles form the candidate set for downstream ranking.

\subsection{Evaluation}
We evaluate our method's performance on traditional and normative diversity metrics \citep{vrijenhoekRADioIntroductionMeasuring2024}. Accuracy and ranking metrics are not used because our approach is designed solely for candidate generation, not ranking, which is the function of the downstream NRS.

\subsubsection{Relevance Evaluation Metrics}
We employ established relevance evaluation metrics to assess both the position of the first relevant item and the overall ranking discrimination.

\noindent\textbf{Mean Reciprocal Rank (MRR):}
We measure ranking quality using Mean Reciprocal Rank, which quantifies how early the first relevant item appears. Given a ranked list of candidate items and a set of relevant items $\mathcal{R}$, MRR is defined as
\begin{equation}
    \text{MRR} = \frac{1}{|Q|}\sum_{q=1}^{|Q|} \frac{1}{\min_{i \in \mathcal{R}_q} \text{rank}(i)}
\end{equation}
where $\text{rank}(i)$ denotes the position of item $i$ in the ranked list. When multiple queries are evaluated, MRR is averaged across queries. Intuitively, MRR rewards systems for retrieving at least one relevant item early, with higher values indicating better ranking.

\noindent\textbf{Area Under the ROC Curve (AUC):}
We measure ranking discrimination using AUC. Given a ranked list with $m$ relevant items $\mathcal{R}$ and $p$ irrelevant items, AUC is computed as
\begin{equation}\text{AUC} = \frac{U}{m \times p}, \quad \text{where} \quad U = \sum_{i \in \mathcal{R}} \text{rank}(i) - \frac{m(m+1)}{2}\end{equation}
This metric measures the probability that a randomly selected relevant item outranks a random irrelevant item. 

\subsubsection{Diversity Evaluation Metrics}
\label{sec:metrics-classic}
In contrast to relevance metrics, which focus on ranking correctness, diversity metrics capture how broadly, unexpectedly, and representatively content is exposed.

\noindent\textbf{Intra-List Diversity:}  
We measure intra-list diversity using average pairwise dissimilarity.
Let $\mathbf{f}_i$ and $\mathbf{f}_j$ denote feature vectors of candidate articles, and $\mathcal{P}$ be the set of all article pairs. Then
$D_{\text{avg}} = \frac{1}{|\mathcal{P}|} \sum_{(i,j) \in \mathcal{P}} \left[ 1 - \cos(\mathbf{f}_i, \mathbf{f}_j) \right]$.

\noindent\textbf{Inverse Calibration:}  
Inverse Calibration measures how well the candidate set distribution aligns with a user's historical preferences (higher values = better alignment). Using the notation from \citet{vrijenhoekRADioIntroductionMeasuring2024}, it is computed as
\begin{equation}\text{ICal} = \sum_c Q^*(c \mid R) \left[ 1 - \text{JSD}\big(P^*(c \mid H), Q^*(c \mid R)\big) \right]\end{equation}
where $c$ denotes a news category, $P^*(c \mid H)$ and $Q^*(c \mid R)$ are the historical and candidate distributions over categories.

\noindent\textbf{Serendipity:}  
Serendipity measures the average dissimilarity of the candidate sets from a user's past clicks, using a continuous measure of similarity, namely cosine. It is computed as
$S = \frac{1}{|\mathcal{C}|} \sum_{c \in \mathcal{C}} \left[ 1 - \max_{h \in \mathcal{H}} \cos(\mathbf{f}_c, \mathbf{f}_h) \right]$,
where $\mathcal{C}$ is the candidate set and $\mathcal{H}$ is the set of article in the user history.

\noindent\textbf{Novelty:}  
Novelty is a simplified, binary version of serendipity. It counts the proportion of candidates that are ``novel'' based on a strict threshold. An article is considered novel if its maximum similarity to any previously clicked item is below a certain threshold $\theta$:
$\text{Novelty} = \left|\left\{ c \in \mathcal{R} \,\middle|\, \max_{h \in \mathcal{H}}\cos(\mathbf{f}_c, \mathbf{f}_h) < \theta \right\}\right| / |\mathcal{C}|$,
where we set $\theta = 0.65$.

\noindent\textbf{Similarity:} Similarity measures the average similarity between candidate sets and user histories: 
\begin{equation} \text{Similarity} = \frac{1}{|I| \cdot |J|} \sum_{i=1}^{|I|} \sum_{j=1}^{|J|} \frac{\vec{c}_i \cdot \vec{h}_j}{\|\vec{c}_i\| \cdot \|\vec{h}_j\|} \end{equation} 
where $\vec{c}_i$ and $\vec{h}_j$ are the components of aspect vectors of candidate set and the user history, respectively.

\subsubsection{Democratic Models}
\label{sec:metrics-democracy}
The evaluation of \texttt{A2CG} employs four democratic models \citep{karppinenUsesDemocraticTheory2012, helbergerDemocraticRoleNews2019}: 
\begin{enumerate}
    \item \textbf{The Liberal model} prioritizes a user's personal preferences and autonomy.
    \item \textbf{The Participatory model} focuses on the common good to help users be active citizens.
    \item \textbf{The Deliberative model} fosters debate by neutrally presenting diverse viewpoints.
    \item \textbf{The Critical model} challenges the status quo by amplifying unheard voices to inspire action against injustice.
\end{enumerate}

To quantitatively assess them, we operationalize the qualitative model \citep{vrijenhoek2021recommenders,vrijenhoekRADioIntroductionMeasuring2024} (see Appendix \ref{app:radio}) into a continuous scoring function. The higher the score for each model, the greater the degree of alignment of the candidate generator model with the democratic model. We define the score as 
\begin{equation}
Score = \left( \sum_{i=1}^{n} (1 - ABS(\text{Method Value}_i - \text{Target Value}_i)) \right)  / n   
\end{equation}
where $n$ is the number of metrics (Calibration etc.) associated with the democratic model. 
We set the target values as follows: \textbf{Low} = 0, \textbf{Medium} = 0.5, and \textbf{High} = 1. The target values are indicated in Table \ref{tab:operationalization}. For the computation of Deliberative, we measure the Representation divergence between the uniform distribution and the candidates. 

\begin{table}[t!b]
\centering
\small
\caption{Operationalization of Normative Democratic Models. Qualitative scores are taken from \citep{vrijenhoekRADioIntroductionMeasuring2024}, $t$ = target value.}
\label{tab:operationalization}
\setlength{\tabcolsep}{2pt}
\begin{tabular}{lcccc}
\toprule
\textbf{Metric} & \textbf{Liberal} & \textbf{Participatory} & \textbf{Deliberative} & \textbf{Critical} \\
\midrule
\textbf{Calibration} & Low ($t$=0) & High ($t$=1) & - & - \\
\textbf{Fragmentation} & High ($t$=1) & Low ($t$=0) & Low ($t$=0) & - \\
\textbf{Activation} & - & Medium ($t$=0.5) & Low ($t$=0) & High ($t$=1) \\
\textbf{Representation} & - & Reflective ($t$=0) & Equal ($t$=0) & Inverse  ($t$=1) \\
\textbf{Altern. Voices} & - & Medium ($t$=0.5) & - & High ($t$=1) \\
\bottomrule
\end{tabular}
\end{table}

\subsubsection{Baselines}
Our experimental evaluation includes a set of candidate generators designed to benchmark the effectiveness of our \texttt{A2CG} framework. These baselines represent common strategies for candidate generation, ranging from simple random selection to more sophisticated methods presented in this order.

\noindent\textbf{Uniform Random:} This baseline selects a set of candidates completely at random from the entire available articles, without considering user history or article aspects. It serves as a control to measure the baseline level of diversity and relevance, as it has no explicit mechanism for either.

\noindent\textbf{History Average:} This relevance-based baseline calculates a single query vector by averaging the embedding vectors of all articles in a user's reading history. It then uses this average vector to perform a nearest-neighbor search in the vector database to retrieve the most similar articles. This method prioritizes relevance to a user's overall past interests but has no built-in mechanism for diversity. 

\noindent\textbf{K-Means Random:} This baseline aims for a balance between diversity and relevance by first clustering (50 clusters) the entire article corpus using K-Means and then randomly selecting one article from each cluster to form a diverse candidate set. This ensures that the selected candidates span the major aspects present in the data. 

\noindent\textbf{K-Means Greedy:} This diversity-focused baseline is an iterative algorithm that builds a diverse set of candidates within a K-Means cluster (50 clusters). It starts by picking one random article. In each subsequent step, it calculates the dissimilarity of all other articles to the current set of chosen candidates and selects the next article based on a probability distribution weighted by this dissimilarity. This ensures that each article adds a new perspective to the candidate set. 

\noindent\textbf{Dataset Candidate Generator:} Since our model intervenes at the candidate generation stage, it makes sense to compare it against the candidates precomputed for the given dataset. We work with the \texttt{MIND} dataset (see below) and consider the \texttt{MIND} behavior logs impressions  as the output of its proprietary candidate generation algorithm. To differentiate between the \texttt{MIND} dataset and its candidate generation, we will use the term \textit{the \texttt{MIND} baseline} to indicate its algorithm for generating impressions.

\section{Experimental Setup}
\subsection{Dataset}
The Microsoft News Dataset (\texttt{MIND}) \citep{wuMINDLargescaleDataset2020a} is a widely used, large-scale news corpus for NRS. While other news datasets like Adressa \citep{gullaAdressaDatasetNews2017} and EB-NERD \citep{kruseEBNeRDLargeScaleDataset2024a} are available, they are typically single-publisher in nature, sourced directly from individual news organizations. In contrast, MIND stands out as the only large-scale aggregate portal news dataset, as it is compiled from the Microsoft News aggregator. This makes it uniquely suited for our candidate generation framework, which requires a diverse, multi-source pool of articles to simulate a global retrieval environment. 

For our experiments, we used the \texttt{MIND}-large version, with a more extensive set of user sessions and articles. We followed general practices \citep{ianaTrainOnceUse2024},  re-splitting the dataset, fine-tuning our model on the validation set, and evaluating  on the test set. Since the original dataset is pre-filtered, we treated all 9,107/6,997 articles from the validation/test set as the global candidate pool. This mimics a real-world scenario where the goal is to narrow a large daily volume of news down to a relevant subset. We set the retrieval target to 25 items, aligning with the average number of impressions per user in the training data.

We augment the \texttt{MIND} dataset with additional aspects: sentiment, political leaning, and frames. We use pre-trained classifiers with title and abstract text as input. Each classifier targets a distinct construct (sentiment, political leaning, and media framing), which differ in their label spaces. We therefore use published task-specific models or established practice in prior work, rather than a single unified classifier that would compromise performance and validity. These augmented labels are treated as ground truth for training the aspect-specific embedding functions in our framework (\S~\ref{subsec:multi-aspects-embedding}).

\noindent\textbf{Sentiment:} We use a fine-tuned DistilBERT model \citep{sanhDistilBERTDistilledVersion2020, tabularisai_2025}, trained on a sentiment analysis corpus, to classify each article into  five  categories (\{Very\} Negative, Neutral,\{Very\} Positive). The reported F1 score of the classifier is 0.93 \citep{tabularisai_2025}.

\noindent\textbf{Political Leaning:} We employ a two-stage process to classify the political leaning of articles. First, we use the DEBATE model \citep{burnham2024political} to classify articles as either political or non-political. For articles identified as political, we then use the \texttt{LLAMA3.1-70B} model \citep{llama3modelcard} to further classify their leaning as left, center, right, or neutral. Our approach follows \citet{ceron2025politicalcontentllmspre}, where the reported F1 score of the classifier is 0.76.

\noindent\textbf{Media Frames:} Frames highlight or emphasize certain aspects of a debate, providing a rich representation of perspective in news articles \cite{entman1993framing}. We use a fine-tuned XLM-RoBERTa-based framing classifier \citep{daffara2025generalizability, DBLP:journals/corr/abs-1911-02116} to identify the dominant media frames present in each article (e.g., economic consequences, human interest, morality). This enables us to capture a more nuanced form of diversity beyond simple topical variation. The reported F1 score of the classifier is 0.68.

While the individual classifiers vary in performance, \texttt{A2CG} is architecturally robust to classification noise. The decoder is jointly trained over all aspects simultaneously, reducing sensitivity to errors in any single classifier. Furthermore, the VQ-VAE quantization clusters similar embeddings into shared codebook entries, smoothing out minor misclassifications before they propagate to the retrieval stage.

\subsection{Implementation Details}
\subsubsection{Aspect-Specific Embeddings}
We use a ModernBERT-base model from HuggingFace as encoder for our Sentence Transformer. We fine-tune four separate models, each corresponding to one of the four aspects: sentiment, political leaning, category (from \texttt{MIND}), and frames. The supervised contrastive learning objective is optimized using the AdamW optimizer. We train each model until it converges. The temperature parameter $\tau$ for the contrastive loss is set to 0.1.

\subsubsection{Article Quantization with VQ-VAE}
For each aspect, we train an independent VQ-VAE. The encoder and decoder networks are implemented as simple feed-forward networks. The input dimension for each VQ-VAE is the embedding size of the Sentence Transformer (d=768). We use Optuna to optimize the hyperparameters for each VQ-VAE model.

\subsubsection{Decoder-Only Modeling}
We utilize a decoder-only transformer model based on the Llama 3 architecture \citep{llama3modelcard,touvronLLaMAOpenEfficient2023a}, which we train from scratch. (For the full formulation of the Transformer's architecture, please refer to \citet{vaswaniAttentionAllYou2017a}.)
The input to this model is the concatenated sequence of quantized aspect codes. The training objective is the cross-entropy loss over the predicted next code in the sequence.

\subsubsection{Retrieval and Evaluation}
For candidate retrieval, we use Milvus, a high-dimensional vector database, to store the concatenated embeddings of all articles in the \texttt{MIND} test set. To improve search speed, we reduced the dimensionality of each aspect to 128 using Singular Value Decomposition (SVD). We then used the Hierarchical Navigable Small World (HNSW) algorithm to perform an efficient nearest-neighbor search. To get 50 candidate articles, we retrieve the two most similar articles for each of the top 25 predictions from the beam search (beam width 25).

\subsubsection{Evaluation Protocol}
We use the metrics detailed in Sections \ref{sec:metrics-democracy} and \ref{sec:metrics-classic} to assess the diversity of the generated candidate sets. We compare our approach to the baseline candidate generation mechanism of the \texttt{MIND} dataset, which is not explicitly designed for diversity. Our analysis will specifically focus on how \texttt{A2CG}, with different aspect-biasing configurations, affects these diversity metrics and their trade-off with relevance.

\section{Results}

\noindent\textbf{RQ1: To what extent does \texttt{A2CG} enable parameterized intervention over democratic goals?}

\begin{figure*}
    \centering
    \includegraphics[width=0.8\linewidth]{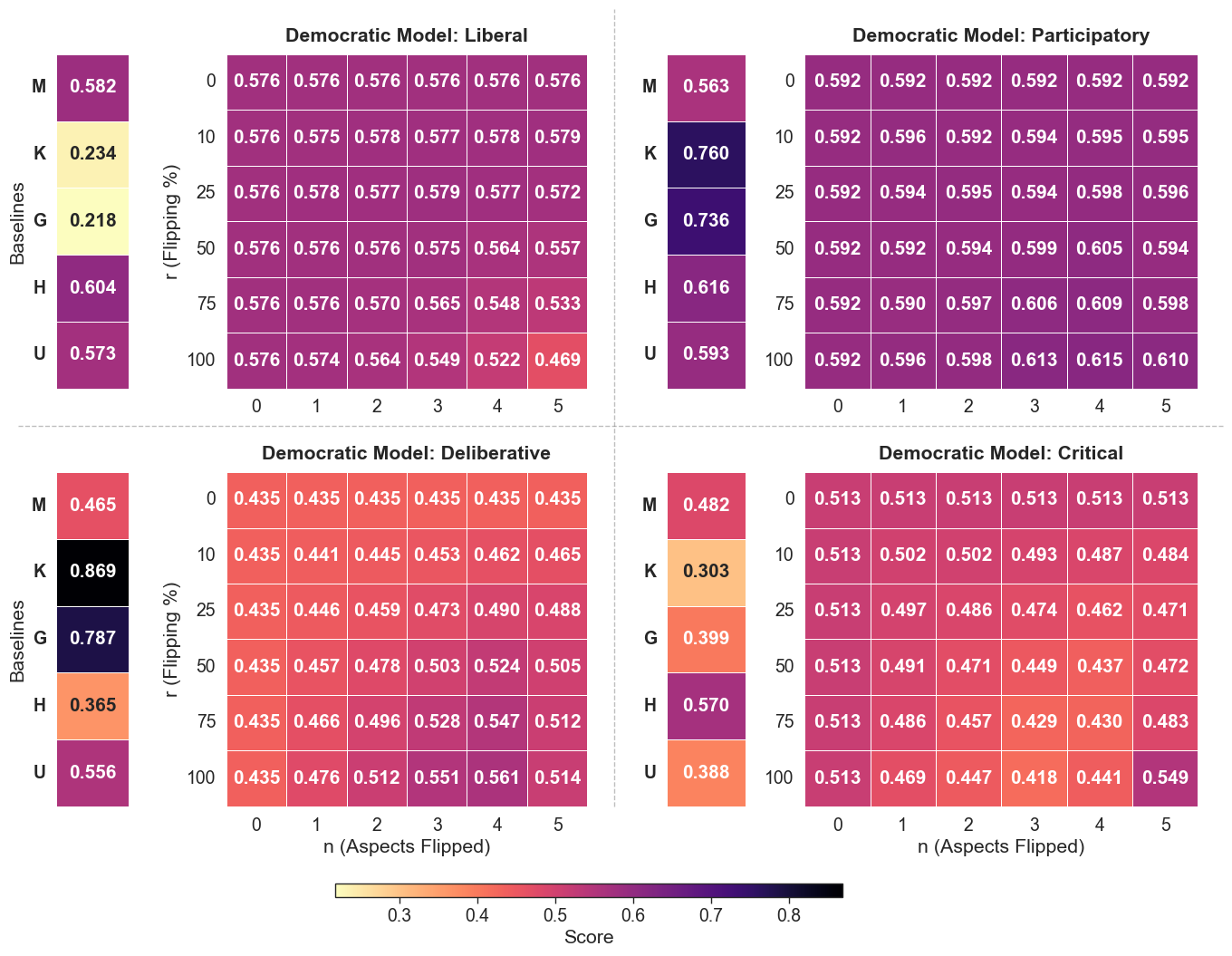}
    \caption{\texttt{A2CG} performance across flipping parameters ($n, r$) relative to baselines. The left strip in each subplot displays baseline scores (M: MIND, K: K-Means Random, G: K-Means Greedy, H: History Average, U: Uniform Random), while the square heatmaps to the right show \texttt{A2CG} scores as a function of aspects flipped ($n$) and flipping chance ($r$). Darker colors represent higher scores. All configurations are statistically significant ($p<0.0001$, Wilcoxon signed-rank test), with the exception of $n=1$, $r=0.75$. See full statistics in Appendix \ref{app:significant}.}
    \label{fig:four_models}
\end{figure*}

To evaluate the extent to which A2CG enables parameterized intervention, we analyze the relationship between the intervention parameters ($n$ and $r$) and the resulting alignment with the four democratic models: \textit{Liberal}, \textit{Participatory}, \textit{Deliberative}, and \textit{Critical}. 
Figure \ref{fig:four_models} illustrates \texttt{A2CG}'s performance across these models and five baselines. The \textit{score} represents the degree of alignment with the  normative goals of each democratic model on a single run. The \texttt{A2CG} scores are a function of $n$ and $r$, where the top left ($n$=0, $r$=0) cell in each subchart of Figure \ref{fig:four_models} denotes the unintervened \texttt{A2CG}. 

\paragraph{Baseline}
With the exception of MIND, each baseline method exhibits a fixed structural bias that constrains its democratic alignment, highlighting \texttt{A2CG}'s key contribution of controllability over pointwise superiority. The History Average method, which focuses purely on a user's history, excels at the \textit{Liberal} model (0.604), which prioritizes individual preference. It also, perhaps counter-intuitively, scores highest on the \textit{Critical} model (0.570). This suggests that by averaging a user's history, the resulting candidates are generic enough to challenge the specific dominant narratives of any single article. Conversely, it is the lowest on \textit{Deliberative} model (0.365), which emphasizes reasoned debate and exposure to diverse perspectives. This indicates that simply averaging a user's historical preferences does little to foster engagement with opposing viewpoints or promote thoughtful deliberation across different frames. 

In contrast, random or cluster-based baselines, e.g., K-Means Random, achieve higher scores on \textit{Participatory} (0.760) and \textit{Deliberative} (0.869) models, reflecting their greater ability to introduce novel content and encourage collective engagement, at the cost of reduced alignment with individual preferences (\textit{Liberal}).

Our \texttt{A2CG} framework, however, can be flexibly tuned to prioritize any of the four democratic models. This adaptability addresses the core limitation of the baselines. For the \textit{Liberal} model, a moderate configuration ($n$=3, $r$=25\%) achieves a score of 0.579, comparable to the strong \texttt{MIND} baselines. Predictably, the most aggressive setting ($n$=5, $r$=100\%) has the lowest score on this metric, confirming that excessive diversity injection undermines individual preference. 

To align with \textit{Participatory} and \textit{Deliberative} models, more aggressive flipping is necessary. The $n$=4 and $r$=100\% maximize the score (Participatory 0.615 and Deliberative 0.561), comparable to the \texttt{MIND} baseline. The \textit{Liberal} score does not lower much (0.522), indicating \texttt{A2CG} can broaden perspectives without completely abandoning personalization.
For the \textit{Critical} model, which challenges dominant narratives, the most aggressive configuration ($n$=5, $r$=100\%) is the most effective. It achieves the highest scores among \texttt{A2CG} variants (0.549), higher than all baselines except  History Average.

Our key insights are: (1) \texttt{A2CG} configurations can selectively strengthen specific democratic functions, and (2) \texttt{A2CG}'s architectural flexibility allows us to target a normative \textit{sweet spot}, i.e., trade-off between various democratic models.

\hfill\\
\noindent\textbf{RQ2: How do different democratic configurations of \texttt{A2CG} reshape exposure for distinct user archetypes under the four democratic models?}

\begin{figure*}[tb!]
     \centering
     % First Subfigure: The Heatmaps (Democratic Models)
     \begin{subfigure}[t]{\textwidth}
        \centering
        \includegraphics[width=\linewidth]{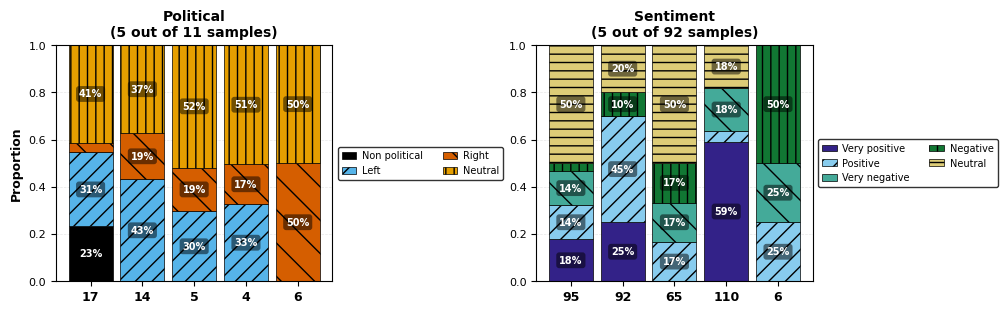}
        \caption{User-Content political alignment and sentiment polarity archetypes (five samples each).}
        \label{fig:archetypes}
     \end{subfigure}
     \hfill
     % Second Subfigure: The Archetypes (Political Exposure)
     \begin{subfigure}[c]{\textwidth}
        \centering
        \includegraphics[width=\linewidth]{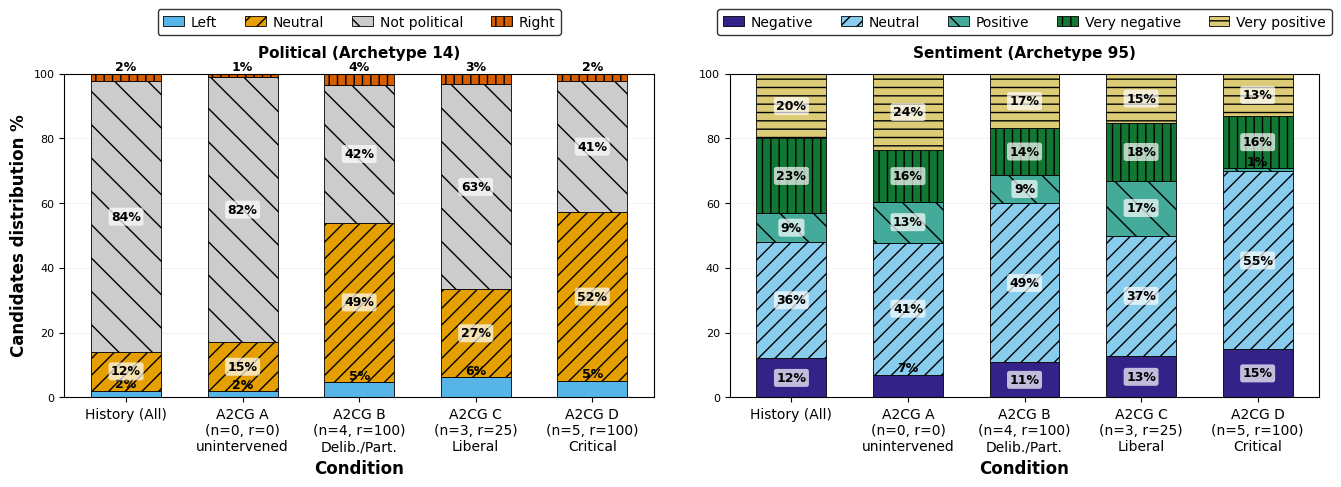}
        \caption{\texttt{A2CG}'s Impact on political alignment archetype 14 and sentiment polarity archetype 95.}
        \label{fig:archetypes_distributions}
    \end{subfigure}
     \caption{(a) archetype examples; (b)  example of  political and sentiment archetypes for user-specific impact.}
     \label{fig:combined_results}
\end{figure*}
The previous RQ demonstrates that its tunable intervention mechanism allows for democratic model-specific targeting, effectively addressing the ``one-size-fits-all'' limitation of the baselines. However, statistical averages do not fully reveal how individual users experience these interventions. To demonstrate the impact, we derive user-content archetypes from the aspect indices used by the \texttt{A2CG} decoder. These archetypes provide a  snapshot of a user's latent interests, i.e., the specific content perspectives they are likely to consume based on their unique reading history. 

Figure \ref{fig:archetypes} illustrates the distinct latent archetypes derived from predicted aspect indices. The political archetypes range from highly partisan (Political Archetype 6) to diverse (Political Archetype 17). The sentiment archetypes range from predominantly positive (Sentiment Archetype 110) to predominantly negative (Sentiment Archetype 6).  By identifying distinct user archetypes based on their histories, which ranges from politically active users to those who consume primarily non-political content, we can observe how different democratic configurations reshape candidates' content. 

Figure \ref{fig:archetypes_distributions} demonstrates the impact of these configurations on two specific profiles: Political Archetype 14 and Sentiment Archetype 95. In both archetypes, the unintervened state (\texttt{A2CG} A), the distribution closely follows the user's history. However, \texttt{A2CG} allows for significant shifts toward different democratic ideals.

\textit{Deliberative/Participatory} (\texttt{A2CG} B): Under aggressive flipping ($n=4, r=100$), in both archetypes, the model systematically shifts distributions toward neutrality, but through different mechanisms across domains: in political contexts, neutrality replaces non-political content (indicating viewpoint injection), accompanied by slight increases in right-leaning (1\% to 4\%) and left-leaning (2\% to 5\%) articles; whereas in sentiment contexts, neutrality primarily compresses extreme polarity (indicating intensity suppression). The effect is substantially stronger in political distributions (15\% to 49\%), with a more moderate shift in sentiment distributions (41\% to 49\%). This aligns with deliberative/participatory characteristics in that the model increases exposure to diverse viewpoints and reduces extreme polarity. 

\textit{Liberal} (\texttt{A2CG} C): A moderate intervention ($n=3, r=25$) preserves personalization in both archetypes, but with noticeable shifting. In the political context, non-political content remains dominant (63\%), yet is substantially reduced compared to user history, alongside an increase in neutral political content (27\%), indicating some injection of political material. In the sentiment context, the distribution does not closely follow user history; instead, strong positive sentiment is reduced while moderate categories increase, resulting in a more balanced but less polarized distribution (neutral at 37\%). Compared to deliberative/participatory, the impact is smaller and maintains closer alignment with user history, but still nudges toward reduced polarization.

\textit{Critical} (\texttt{A2CG} D): In the most aggressive configuration ($n=5, r=100$), the model substantially reshapes both political and sentiment distributions by increasing expressive and opinionated content. At the political level, non-political content is reduced (41\%), while both left-leaning (5\%) and neutral political content (52\%) increase, indicating a strong shift toward politically engaged material, effectively surfacing alternative perspectives that challenge the user’s historical status quo. At the sentiment level, neutrality becomes dominant (55\%), while both extreme positive (13\%) and negative (16\%) sentiment are reduced. In sum, this  intervention surfaces diverse perspectives and reduces  polarization.

\hfill\\
These changes demonstrate how \texttt{A2CG} offers the user fresh perspectives (the extra neutral, left and right articles), which is only possible when the entire process is carried out in the candidate generation stage rather than by re-ranking. This confirms that diversity injection meaningfully shifts individual content exposure toward the targeted democratic ideal, beyond what aggregate metrics alone reveal.
\begin{table*}[tb!]
    \small
    \centering
    \caption{The performance of each model based on personalization and diversity metrics. Numbers with symbols ($\bot$, $\top$) are bolded to indicate the lowest and highest scores. The \textcolor{ForestGreen}{$\uparrow$} denotes that the \texttt{A2CG} number is higher than the previous row, \textcolor{red}{$\downarrow$} otherwise. For all metrics, higher values are better. All configurations are statistically significant ($p<0.0001$, Wilcoxon signed-rank test).}
    \label{tab:diversity_only}
    \begin{tabular}{lccccccc}
        \toprule
        \textbf{Method} &
        \multicolumn{4}{c}{\textbf{Personalization}} &
        \multicolumn{3}{c}{\textbf{Diversity}} \\
        \cmidrule(lr){2-5} \cmidrule(lr){6-8}
        & \textbf{AUC} & \textbf{MRR} & \textbf{Inv Calibration} & \textbf{Similarity} 
        & \textbf{Intra-List Div.} & \textbf{Serendipity} & \textbf{Novelty} \\
        
        \midrule
        \textbf{\texttt{MIND} baseline} & \textbf{0.4996} $\top$ & \textbf{0.2638} $\top$ & 0.4333 & 0.3709 & 0.6397 & 0.3426 & 0.4177 \\
        \textbf{K-Means Random} & \textbf{0.0010} $\bot$ & 0.0057 & 0.4671 & \textbf{0.3820} $\top$ & 0.6008 & \textbf{0.3306} $\bot$ & \textbf{0.3818} $\bot$ \\
        \textbf{K-Means Greedy} & \textbf{0.0010} $\bot$ & 0.0016 & 0.4361 & 0.3534 & 0.6472 & 0.3755 & 0.5034 \\
        \textbf{History Avg.} & 0.0089 & 0.0420 & \textbf{0.5102} $\top$ & 0.3085 & \textbf{0.4167} $\bot$ & 0.3687 & 0.5165 \\
        \textbf{Uniform Random} & 0.0110 & 0.0021 & 0.4379 & 0.3695 & 0.6200 & 0.3560 & 0.4583 \\
        \hline\hline
        \textbf{\texttt{A2CG} (Vanilla)} & 0.0124 & \textbf{0.0011} $\bot$ & \textbf{0.4078} $\bot$ & 0.3716 & 0.5643  & 0.3650 & 0.4740 \\
        \rowcolor{lightgray}
        \multicolumn{8}{c}{n=1} \\
        \textbf{- \textbackslash w 10\%} & 0.0011 & 0.0124 & 0.4102 & 0.3687 & 0.5731 & 0.3670 & 0.4784 \\
        \textbf{- \textbackslash w 50\%} & 0.0134 \textcolor{ForestGreen}{$\uparrow$} & 0.0012 \textcolor{red}{$\downarrow$} & 0.4180 \textcolor{ForestGreen}{$\uparrow$} & 0.3579 \textcolor{red}{$\downarrow$} & 0.6030 \textcolor{ForestGreen}{$\uparrow$} & 0.3745 \textcolor{ForestGreen}{$\uparrow$} & 0.4950 \textcolor{ForestGreen}{$\uparrow$} \\
        \textbf{- \textbackslash w 100\%} & 0.0152 \textcolor{ForestGreen}{$\uparrow$} & 0.0012 \textcolor{red}{$\downarrow$} & 0.4247 \textcolor{ForestGreen}{$\uparrow$} & 0.3449 \textcolor{red}{$\downarrow$} & 0.6337 \textcolor{ForestGreen}{$\uparrow$} & 0.3839 \textcolor{ForestGreen}{$\uparrow$} & 0.5161 \textcolor{ForestGreen}{$\uparrow$} \\
        \rowcolor{lightgray}
        \multicolumn{8}{c}{n=3} \\
        \textbf{- \textbackslash w 10\%} & 0.0011 & 0.0129 & 0.4125 & 0.3616 & 0.5933 & 0.3720 & 0.4898 \\
        \textbf{- \textbackslash w 50\%} & 0.0145 \textcolor{ForestGreen}{$\uparrow$} & 0.0014 \textcolor{red}{$\downarrow$} & 0.4260 \textcolor{ForestGreen}{$\uparrow$} & 0.3262 \textcolor{red}{$\downarrow$} & 0.6667 \textcolor{ForestGreen}{$\uparrow$} & 0.3970 \textcolor{ForestGreen}{$\uparrow$} & 0.5470 \textcolor{ForestGreen}{$\uparrow$} \\
        \textbf{- \textbackslash w 100\%} & 0.0163 \textcolor{ForestGreen}{$\uparrow$} & 0.0021 \textcolor{red}{$\downarrow$} & 0.4352 \textcolor{ForestGreen}{$\uparrow$} & 0.2823 \textcolor{red}{$\downarrow$} & \textbf{0.6964} $\top$ & 0.4285 \textcolor{ForestGreen}{$\uparrow$} & 0.6201 \textcolor{ForestGreen}{$\uparrow$} \\
        \rowcolor{lightgray}
        \multicolumn{8}{c}{n=5} \\
        \textbf{- \textbackslash w 10\%} & 0.0011 & 0.0113 & 0.4148 & 0.3545 & 0.6112 & 0.3738 & 0.4899 \\
        \textbf{- \textbackslash w 50\%} & 0.0085 \textcolor{ForestGreen}{$\uparrow$} & 0.0012 \textcolor{red}{$\downarrow$} & 0.4311 \textcolor{ForestGreen}{$\uparrow$} & 0.3025 \textcolor{red}{$\downarrow$} & 0.6872 \textcolor{ForestGreen}{$\uparrow$} & 0.4010 \textcolor{ForestGreen}{$\uparrow$} & 0.5390 \textcolor{ForestGreen}{$\uparrow$} \\
        \textbf{- \textbackslash w 100\%} & 0.0054 \textcolor{red}{$\downarrow$} & 0.0012 \textcolor{red}{$\downarrow$} & 0.4368 \textcolor{ForestGreen}{$\uparrow$} & \textbf{0.2129}$\bot$ \textcolor{red}{$\downarrow$} & 0.5904 \textcolor{red}{$\downarrow$} & \textbf{0.4492} $\top$ & \textbf{0.6294} $\top$ \\
        \bottomrule
    \end{tabular}
\end{table*}

\hfill\\
\noindent\textbf{RQ3: How does \texttt{A2CG}'s multi-aspect approach balance the trade-off between diversity and personalization?}

We now analyze how the \texttt{A2CG} framework manages the inherent tension between recommendation relevance and various forms of content diversity. Table \ref{tab:diversity_only} shows personalization (relevance) and diversity metrics of our methods and baselines. Full table of results can be seen in Appendix \ref{app:full_table}.

\noindent\textbf{MIND's algorithm.} The \texttt{MIND} baseline achieves the highest AUC (0.4996) and MRR (0.2638) among all methods. This is expected due to the typical construction method of news recommendation dataset: users are only presented with articles retrieved by the candidate selection algorithm; consequently, no relevance judgments are collected for all other articles. Since our \texttt{A2CG} framework intervenes at the candidate generation stage, most of our predicted articles fall into this category. Thus, the much higher ARR and MRR numbers for the \texttt{MIND} baseline represent exposure bias rather than a demonstrable superiority of the MIND model. \textit{Inverse Calibration} is lower than other baselines, showing that, without using clicks, the recommendations are less personalized and match the user's history worse than other baselines.

\noindent\textbf{Personalization \& diversity tradeoff.} \textit{MRR}, and \textit{Similarity} tend to decrease as diversity-focused strategies become stronger. For example, \textit{K-Means}  and highly-intervening \texttt{A2CG} configurations improve \textit{Intra-List Diversity}, \textit{Serendipity}, and \textit{Novelty}, but often at the cost of lower \textit{MRR} or slightly worse \textit{Inverse Calibration}. This reflects an inherent trade-off:  more varied content reduces the algorithm’s ability to perfectly match the user’s top historical preferences.

\noindent\textbf{AUC \& inverse calibration.} \textit{AUC} and \textit{Inverse Calibration} follows the trend of diversity metrics. This happens because increasing diversity can spread relevant items more evenly throughout the ranking, which can raise \textit{AUC} and \textit{Inverse Calibration} even if the very top predictions are slightly less accurate. These findings are consistent with \citet{silva2023representation}, who observe that relevance can improve alongside diversity metrics.

\noindent\textbf{\texttt{A2CG}.} Unlike baselines with fixed structural biases, \texttt{A2CG}'s explicit intervention parameters allow continuous calibration between personalization and diversity without retraining. As the flipping increases, diversity metrics such as \textit{Intra-List Diversity}, \textit{Serendipity}, and \textit{Novelty} consistently rise, showing that the model effectively explores broader aspects beyond the user’s immediate history. At the same time, personalization metrics like \textit{Similarity} and \textit{MRR} decrease moderately, reflecting the expected trade-off when introducing more varied items. These patterns highlight that \texttt{A2CG} can flexibly adjust the trade-off between relevance and diversity, providing a tunable framework for generating both personalized and diverse recommendations.

\section{Conclusion and Future Work}
In this paper, we propose Aspect-Aware Candidate Generation (\texttt{A2CG}) to democratize news recommender systems by diversifying news recommendation while accounting for user reading preferences within democratic models. Our evaluation addresses three key research questions, demonstrating \texttt{A2CG}'s effectiveness and flexibility. Our findings show that \texttt{A2CG} performs competitively against both baselines and the proprietary \texttt{MIND} algorithm. Furthermore, by integrating a mechanism for diversity injection, \texttt{A2CG} becomes a versatile and tunable solution for balancing personalization with normative diversity. Finally, our analysis of the user archetypes indicates that \texttt{A2CG} can introduce fresh perspectives aligned with targeted democratic models, providing additional empirical evidence that improving normative diversity is successfully achieved at the candidate generation stage. For future work, we plan to integrate \texttt{A2CG} into a comprehensive simulation model. This simulation environment will provide a controlled and efficient way to test and optimize NRS, offering insights that would be difficult to obtain through real-world deployment.

\section{Limitation}
Democratic ideals are philosophically rich concepts that cannot be fully captured by any finite set of metrics. Our contribution is therefore not a complete formalization of democratic theory, but a practical operational bridge between normative models and candidate-generation mechanisms via the \texttt{RADio*} framework. Moreover, 
our evaluation is entirely offline and conducted on a single dataset, \texttt{MIND}. This limits the generalizability of our findings to user populations and news ecosystems. While \texttt{MIND} is one of the few large-scale, multi-publisher datasets available for evaluating candidate generation, and its aggregator setting provides a heterogeneous article pool where diversity interventions are particularly relevant, it may not fully capture the dynamics of real-world news platforms. Furthermore, although our algorithm explicitly promotes exposure to diverse perspectives, users may still selectively engage with content that aligns with their existing beliefs. This makes it difficult to assess the real-world impact of our approach. Prior work suggests that ranking and exposure can influence news consumption behavior \citep{mattis2024nudging}, but the extent to which such effects translate into real changes in user engagement remains uncertain. Overall, while our approach is grounded in theoretical frameworks for diversifying news recommendation is a step toward diversifying perspectives, validating its effectiveness in real-world settings requires future work with online evaluation and user studies.

\section*{Ethical Considerations Statement}

If implemented maliciously, \texttt{A2CG} may prioritize the candidate set of only one perspective, potentially leading to selective exposure of information \citep{hart2009feeling}. Another factor to take into account is the quality of the aspect classifiers. They need constant re-training with updated news coverage given that there are often emerging topics in the media. This ensures that news articles are not misclassified at a high rate. For example, in the context of political leaning, the diversification would be affected and one side of the political spectrum could be overemphasized even though the objective of \texttt{A2CG} is to diversify content. That said, A2CG's architecture mitigates this risk to some extent. Joint decoder training reduces sensitivity to any single classifier, and VQ-VAE codebook smoothing absorbs minor classification errors before retrieval.

At a more general level, when implemented for real-world use cases, the pipeline should ensure it only includes reliable sources to avoid spreading misinformation. 

\section*{Generative AI Usage Statement}

Grammarly was used for correcting grammar errors in the writing process. We have also made use of ChatGPT-5.2 for paraphrasing some sentences in main text of the paper. In addition to that, we also use Copilot for aiding our development process.

\section*{Acknowledgments}
This research has been conducted as part of the \texttt{Multiview} project (\texttt{Multiview} - Künstliche Intelligenz für ausgewogene Nachrichtenempfehlung, P2024-0749), funded by Vector-Stiftung within the MINT-Innovationen 2024 grant program. 

%%
%% The next two lines define the bibliography style to be used, and
%% the bibliography file.
\bibliographystyle{ACM-Reference-Format}
\bibliography{sample-base}

%%
%% If your work has an appendix, this is the place to put it.
\appendix
% \section{Diversity Injection}
% \label{app:diversity_injection}

\FloatBarrier
\section{\texttt{RADio} Metrics}
% \seb{I think there needs to be discussion of why we don't do ANY traditional metrics (notably click prediction)}\hardy{i put it in the intro of the subsection}
\label{app:radio}
We use the \texttt{RADio} framework \citep{vrijenhoekRADioIntroductionMeasuring2024} to derive the democracy, which offers a unified approach to normative diversity metrics
% \ame{Maybe mention that these are the normative metrics?} 
with optional rank-awareness, to obtain operationalization of democracy models. In our setting, we disable the rank-aware component, which applies a positional discount to candidates, as our focus is on diversity irrespective of ranking position. We use category distribution for Calibration and subcategory for Fragmentation, sentiment for Activation, and frame for Representation and Alternative Voices. Metric scores range from 0 (no divergence) to 1 (maximum divergence), without an inherent interpretation as “good” or “bad.”

% \seb{Add one sentence about what \texttt{RADio} does for people who don't know}\hardy{done}
The \texttt{RADio} metrics include:
\begin{itemize}
    \item \textbf{Calibration:} Alignment of topical distribution between candidates and the user's history. Lower values indicate greater similarity.
    \item \textbf{Fragmentation:} Divergence of candidates across users. Lower values indicate more overlap between users.
    \item \textbf{Activation:} Sentiment-based emotional polarization between candidates and the user's history. Lower values indicate a more similar emotional tone.
    \item \textbf{Representation:} Diversity of frames in candidates relative to the entire corpus. Lower values indicate frames are represented proportionally to the dataset. 
    \item \textbf{Alternative Voices:} Diversity of frames in candidates relative to the user's history. Lower values indicate greater similarity to the user's history.
% \tani{I had to read several times to disentangle this definition from Representation. Would Alternative voices be: Assesses the diversity of named entities (e.g., persons) mentioned in news articles relative to the user's history? It contrasts more easily with Representation.} 
\end{itemize}
\section{Traditional \& Normative Results}
\label{app:full_table}
\begin{table*}[h!]
    \small
    \centering
    \caption{Performance on traditional Diversity and \texttt{RADio} Framework Metrics. The \texttt{A2CG} models are stratified into how many aspects are flipped ($n$) and the likelihood of a sample being flipped. Numbers with symbols ($\bot$, $\top$) are bolded to indicate the lowest and highest scores within a group. An asterisk ($\ast$) denotes the overall lowest or highest score. The \textcolor{ForestGreen}{$\uparrow$} denotes that the \texttt{A2CG} number is higher than the previous row, \textcolor{red}{$\downarrow$} otherwise. All configurations are statistically significant ($p<0.001$, Wilcoxon signed-rank test), with the exception of $n=1$, $r=0.75$. See full statistics in Appendix \ref{app:significant}.}
    \label{tab:combined_metrics}
    \resizebox{\linewidth}{!}{%
   \begin{tabular}{l*{9}{c}}
        \toprule
        \textbf{} & \multicolumn{4}{c}{\textbf{Traditional Diversity Metrics}} & \multicolumn{5}{c}{\textbf{RADio Framework Metrics}} \\
        \cmidrule(lr){2-5} \cmidrule(lr){6-10}
        \textbf{Method} & \textbf{Intra-List Div.} & \textbf{Serendipity} & \textbf{Novelty} & \textbf{Similarity} & \textbf{Calibration} & \textbf{Frag.} & \textbf{Act.} & \textbf{Repr.} & \textbf{\makecell{Alt.\\Voices}} \\
        \midrule
        \textbf{\texttt{MIND} baseline} & 0.6397 & 0.3426  & 0.4177 & 0.3709 & \textbf{0.5667} $\top$ & \textbf{0.7316} $\top$& 0.2134 & \textbf{0.6611} $\top$* & \textbf{0.5705} $\top$\\
        \textbf{K-Means Random} & 0.6008 & \textbf{0.3306} $\bot$* & \textbf{0.3818} $\bot$* & \textbf{0.3820} $\top$* & 0.5329 & \textbf{0.0000} $\bot$* &\textbf{ 0.0890} $\bot$* & \textbf{0.3039} $\bot$* &0.5158 \\
        \textbf{K-Means Greedy} & \textbf{0.6472} $\top$ & \textbf{0.3755} $\top$ & 0.5034 & 0.3534 & 0.5639  & \textbf{0.0000} $\bot$* & 0.1551  & 0.4849 & 0.5565 \\
        \textbf{History Avg.} & \textbf{0.4167} $\bot$* & 0.3687  & \textbf{0.5165} $\top$ & \textbf{0.3085} $\bot$ & \textbf{0.4898} $\bot$* & 0.6977 & \textbf{0.6274} $\top$* & 0.5786 & \textbf{0.5044} $\bot$* \\
        \textbf{Uniform Random} & 0.6200 & 0.3560 & 0.4583 & 0.3695 & 0.5621 & 0.7089 & 0.1386 & 0.4853 & 0.5398 \\ 
        \midrule
        \textbf{\texttt{A2CG} (Vanilla)} & \textbf{0.5643} $\bot$ & \textbf{0.3650} $\bot$ & \textbf{0.4740} $\bot$ & \textbf{0.3716} $\top$& 0.5922 & \textbf{0.7433} $\top$* &\textbf{0.3263} $\top$& \textbf{0.6252} $\top$ & 0.5880 \\
        \rowcolor{lightgray}
        \multicolumn{10}{c}{n=1} \\
        \textbf{- \textbackslash w 10\%} & 0.5731 & 0.3670 & 0.4784 & 0.3687 & 0.5898 & 0.7396 & 0.3183 & 0.6200 & \textbf{0.5680} $\bot$ \\
        \textbf{- \textbackslash w 25\%} & 0.5852 \textcolor{ForestGreen}{$\uparrow$}& 0.3699 \textcolor{ForestGreen}{$\uparrow$}& 0.4849 \textcolor{ForestGreen}{$\uparrow$}& 0.3645 \textcolor{red}{$\downarrow$}& 0.5866 \textcolor{red}{$\downarrow$} & 0.7416 \textcolor{ForestGreen}{$\uparrow$} & 0.3079 \textcolor{red}{$\downarrow$} & 0.6122 \textcolor{red}{$\downarrow$} & 0.5722 \textcolor{ForestGreen}{$\uparrow$} \\
        \textbf{- \textbackslash w 50\%} & 0.6030 \textcolor{ForestGreen}{$\uparrow$}& 0.3745 \textcolor{ForestGreen}{$\uparrow$}& 0.4950 \textcolor{ForestGreen}{$\uparrow$}& 0.3579 \textcolor{red}{$\downarrow$}& 0.5820 \textcolor{red}{$\downarrow$}& 0.7348 \textcolor{red}{$\downarrow$} & 0.2929 \textcolor{red}{$\downarrow$} & 0.6010 \textcolor{red}{$\downarrow$} & 0.5782 \textcolor{ForestGreen}{$\uparrow$} \\
        \textbf{- \textbackslash w 75\% } & 0.6190 \textcolor{ForestGreen}{$\uparrow$}& 0.3792 \textcolor{ForestGreen}{$\uparrow$}& 0.5056 \textcolor{ForestGreen}{$\uparrow$}& 0.3514 \textcolor{red}{$\downarrow$}& 0.5785 \textcolor{red}{$\downarrow$} & 0.7306 \textcolor{red}{$\downarrow$} & 0.2801 \textcolor{red}{$\downarrow$} & 0.5908 \textcolor{red}{$\downarrow$} & 0.5862 \textcolor{ForestGreen}{$\uparrow$} \\
        \textbf{- \textbackslash w 100\%} & 0.6337 \textcolor{ForestGreen}{$\uparrow$}& 0.3839 \textcolor{ForestGreen}{$\uparrow$}& 0.5161 \textcolor{ForestGreen}{$\uparrow$}& 0.3449 \textcolor{red}{$\downarrow$}& 0.5753 \textcolor{red}{$\downarrow$} & 0.7233 \textcolor{red}{$\downarrow$} &0.2684 \textcolor{red}{$\downarrow$} & 0.5815 \textcolor{red}{$\downarrow$} & 0.5568 \textcolor{red}{$\downarrow$} \\
        \rowcolor{lightgray}
        \multicolumn{10}{c}{n=2} \\
        \textbf{- \textbackslash w 10\%} & 0.5831 & 0.3698 & 0.4855 & 0.3653 & 0.5875 & 0.7431 & 0.3109 & 0.6123 & 0.5817 \\
        \textbf{- \textbackslash w 25\%} & 0.6069 \textcolor{ForestGreen}{$\uparrow$}& 0.3762 \textcolor{ForestGreen}{$\uparrow$}& 0.4999 \textcolor{ForestGreen}{$\uparrow$}& 0.3564 \textcolor{red}{$\downarrow$}& 0.5816 \textcolor{red}{$\downarrow$} & 0.7349 \textcolor{red}{$\downarrow$} & 0.2918 \textcolor{red}{$\downarrow$} & 0.5974 \textcolor{red}{$\downarrow$} & 0.5681 \textcolor{red}{$\downarrow$}  \\
        \textbf{- \textbackslash w 50\%} & 0.6385 \textcolor{ForestGreen}{$\uparrow$}& 0.3864 \textcolor{ForestGreen}{$\uparrow$}& 0.5235 \textcolor{ForestGreen}{$\uparrow$}& 0.3425 \textcolor{red}{$\downarrow$}& 0.5740 \textcolor{red}{$\downarrow$} & 0.7252 \textcolor{red}{$\downarrow$} & 0.2669 \textcolor{red}{$\downarrow$} & 0.5745 \textcolor{red}{$\downarrow$} & 0.5731 \textcolor{ForestGreen}{$\uparrow$} \\
        \textbf{- \textbackslash w 75\%} & 0.6628 \textcolor{ForestGreen}{$\uparrow$}& 0.3966 \textcolor{ForestGreen}{$\uparrow$}& 0.5472 \textcolor{ForestGreen}{$\uparrow$}& 0.3289 \textcolor{red}{$\downarrow$}& 0.5686 \textcolor{red}{$\downarrow$} & 0.7081 \textcolor{red}{$\downarrow$} & 0.2469 \textcolor{red}{$\downarrow$} & 0.5560 \textcolor{red}{$\downarrow$} & 0.5689 \textcolor{red}{$\downarrow$} \\
        \textbf{- \textbackslash w 100\% } & 0.6818 \textcolor{ForestGreen}{$\uparrow$}& 0.4073 \textcolor{red}{$\downarrow$}& 0.5722 \textcolor{ForestGreen}{$\uparrow$}& 0.3147 \textcolor{red}{$\downarrow$}& 0.5648 \textcolor{red}{$\downarrow$} & 0.6927 \textcolor{red}{$\downarrow$} & 0.2298 \textcolor{red}{$\downarrow$} & 0.5404 \textcolor{red}{$\downarrow$} & 0.5697 \textcolor{ForestGreen}{$\uparrow$} \\
        \rowcolor{lightgray}
        \multicolumn{10}{c}{n=3} \\
        \textbf{- \textbackslash w 10\%} & 0.5933 & 0.3720 & 0.4898 & 0.3616 & 0.5852 & 0.7397 & 0.3031 & 0.5994 & 0.5780 \\
        \textbf{- \textbackslash w 25\%} & 0.6274 \textcolor{ForestGreen}{$\uparrow$}& 0.3818 \textcolor{ForestGreen}{$\uparrow$}& 0.5123 \textcolor{ForestGreen}{$\uparrow$}& 0.3477 \textcolor{red}{$\downarrow$}& 0.5775 \textcolor{red}{$\downarrow$} & 0.7361 \textcolor{red}{$\downarrow$} & 0.2764 \textcolor{red}{$\downarrow$} & 0.5675 \textcolor{red}{$\downarrow$} & 0.5790 \textcolor{ForestGreen}{$\uparrow$} \\
        \textbf{- \textbackslash w 50\% } & 0.6667 \textcolor{ForestGreen}{$\uparrow$}& 0.3970 \textcolor{ForestGreen}{$\uparrow$}& 0.5470 \textcolor{ForestGreen}{$\uparrow$}& 0.3262 \textcolor{red}{$\downarrow$}&0.5689 \textcolor{red}{$\downarrow$} & 0.7193 \textcolor{red}{$\downarrow$} & 0.2447 \textcolor{red}{$\downarrow$} & 0.5275 \textcolor{red}{$\downarrow$} & 0.5736 \textcolor{red}{$\downarrow$} \\
        \textbf{- \textbackslash w 75\%} & 0.6892 \textcolor{ForestGreen}{$\uparrow$}& 0.4121 \textcolor{ForestGreen}{$\uparrow$}& 0.5818 \textcolor{ForestGreen}{$\uparrow$}& 0.3050 \textcolor{red}{$\downarrow$}& 0.5641 \textcolor{red}{$\downarrow$} & 0.6948 \textcolor{red}{$\downarrow$} & 0.2228 \textcolor{red}{$\downarrow$} & 0.4984 \textcolor{red}{$\downarrow$} & 0.5656 \textcolor{red}{$\downarrow$} \\
        \textbf{- \textbackslash w 100\%} & \textbf{0.6964} $\top$*& 0.4285 \textcolor{ForestGreen}{$\uparrow$}& 0.6201 \textcolor{ForestGreen}{$\uparrow$}& 0.2823 \textcolor{red}{$\downarrow$}& \textbf{0.5632} $\bot$ & 0.6616 \textcolor{red}{$\downarrow$} & \textbf{0.2097} $\bot$ &\textbf{0.4768} $\bot$ & 0.5684 \textcolor{ForestGreen}{$\uparrow$} \\
        \rowcolor{lightgray}
        \multicolumn{10}{c}{n=4} \\
        \textbf{- \textbackslash w 10\%} & 0.6038 & 0.3736 & 0.4921 & 0.3576 & 0.5833 & 0.7384 & 0.2953 & 0.5794 & 0.5861 \\
        \textbf{- \textbackslash w 25\%} & 0.6462 \textcolor{ForestGreen}{$\uparrow$}& 0.3860 \textcolor{ForestGreen}{$\uparrow$}& 0.5188 \textcolor{ForestGreen}{$\uparrow$}& 0.3383 \textcolor{red}{$\downarrow$}& 0.5750 \textcolor{red}{$\downarrow$} & 0.7283 \textcolor{red}{$\downarrow$} & 0.2640 \textcolor{red}{$\downarrow$} & 0.5364 \textcolor{red}{$\downarrow$}& 0.5842 \textcolor{red}{$\downarrow$} \\
        \textbf{- \textbackslash w 50\%} & 0.6853 \textcolor{ForestGreen}{$\uparrow$}& 0.4045 \textcolor{ForestGreen}{$\uparrow$}& 0.5581 \textcolor{ForestGreen}{$\uparrow$}& 0.3092 \textcolor{red}{$\downarrow$}& 0.5686 \textcolor{red}{$\downarrow$} & 0.6973 \textcolor{red}{$\downarrow$} & 0.2311 \textcolor{red}{$\downarrow$} & 0.4982 \textcolor{red}{$\downarrow$} & 0.5812 \textcolor{red}{$\downarrow$} \\
        \textbf{- \textbackslash w 75\%} & 0.6942 \textcolor{ForestGreen}{$\uparrow$}& 0.4231 \textcolor{ForestGreen}{$\uparrow$}& 0.5982 \textcolor{ForestGreen}{$\uparrow$}& 0.2799 \textcolor{red}{$\downarrow$}& 0.5691 \textcolor{ForestGreen}{$\uparrow$} & 0.6650 \textcolor{red}{$\downarrow$} & 0.2155 \textcolor{red}{$\downarrow$} & 0.4799 \textcolor{red}{$\downarrow$} & 0.5935 \textcolor{ForestGreen}{$\uparrow$} \\
        \textbf{- \textbackslash w 100} & 0.6676 \textcolor{red}{$\downarrow$}& 0.4447 \textcolor{ForestGreen}{$\uparrow$}& 0.6452 \textcolor{ForestGreen}{$\uparrow$}& 0.2466 \textcolor{red}{$\downarrow$}& 0.5767 \textcolor{ForestGreen}{$\uparrow$}& 0.6211 \textcolor{red}{$\downarrow$} & 0.2224 \textcolor{ForestGreen}{$\uparrow$} & 0.4736 \textcolor{red}{$\downarrow$} & 0.6283 \textcolor{ForestGreen}{$\uparrow$} \\
        \rowcolor{lightgray}
        \multicolumn{10}{c}{n=5} \\
        \textbf{- \textbackslash w 10\%} & 0.6112 & 0.3738 & 0.4899 & 0.3545 & 0.5831 & 0.7401 & 0.2920 & 0.5726 & 0.5873 \\
        \textbf{- \textbackslash w 25\%} & 0.6537 \textcolor{ForestGreen}{$\uparrow$}& 0.3850 \textcolor{ForestGreen}{$\uparrow$}& 0.5103 \textcolor{ForestGreen}{$\uparrow$}& 0.3335 \textcolor{red}{$\downarrow$}& 0.5772 \textcolor{red}{$\downarrow$} & 0.7206 \textcolor{red}{$\downarrow$} & 0.2666 \textcolor{red}{$\downarrow$} & 0.5482 \textcolor{red}{$\downarrow$} & 0.5970 \textcolor{ForestGreen}{$\uparrow$}\\
        \textbf{- \textbackslash w 50\%} & 0.6872 \textcolor{ForestGreen}{$\uparrow$}& 0.4010 \textcolor{ForestGreen}{$\uparrow$}& 0.5390 \textcolor{ForestGreen}{$\uparrow$}& 0.3025 \textcolor{red}{$\downarrow$}& 0.5776 \textcolor{ForestGreen}{$\uparrow$} & 0.6923 \textcolor{red}{$\downarrow$} & 0.2491 \textcolor{red}{$\downarrow$} & 0.5444 \textcolor{red}{$\downarrow$} & 0.6212 \textcolor{ForestGreen}{$\uparrow$} \\
        \textbf{- \textbackslash w 75\%} & 0.6837 \textcolor{red}{$\downarrow$}& 0.4199 \textcolor{ForestGreen}{$\uparrow$}& 0.5740 \textcolor{ForestGreen}{$\uparrow$}& 0.2668 \textcolor{red}{$\downarrow$}& 0.5881 \textcolor{ForestGreen}{$\uparrow$} & 0.6533 \textcolor{red}{$\downarrow$}& 0.2524 \textcolor{ForestGreen}{$\uparrow$} & 0.5582 \textcolor{ForestGreen}{$\uparrow$} & 0.6397 \textcolor{ForestGreen}{$\uparrow$} \\
        \textbf{- \textbackslash w 100\%} & 0.5904 \textcolor{red}{$\downarrow$}& \textbf{0.4492} $\top$*& \textbf{0.6294} $\top$*& \textbf{0.2129} $\bot$* & \textbf{0.6157} $\top$* & \textbf{0.5546} $\bot$ & 0.3188 \textcolor{ForestGreen}{$\uparrow$} & 0.5856 \textcolor{ForestGreen}{$\uparrow$} & \textbf{0.7441} $\top$*\\
        \bottomrule
    \end{tabular}
    }
\end{table*}
\FloatBarrier
\section{Full Statistical Test Results}
\label{app:significant}
\FloatBarrier
% Please add the following required packages to your document preamble:
% \usepackage{multirow}
% Please add the following required packages to your document preamble:
% \usepackage{multirow}
\begin{table}[h!]
\small
\centering
\caption{Wilcoxon signed-rank test results comparing \texttt{A2CG} with no diversity injection against baseline methods on traditional metrics. $W$ denotes the Wilcoxon statistic and $r$ the rank-biserial correlation effect size. All comparisons are statistically significant ($p < 0.0001$). Positive effect sizes indicate \texttt{A2CG} outperforms the baseline; negative values indicate the baseline outperforms \texttt{A2CG}.}
\resizebox{\linewidth}{!}{%
\begin{tabular}{lllllllllllll}
\toprule
\multirow{2}{*}{Baseline Method} & \multicolumn{2}{l}{AUC} & \multicolumn{2}{l}{MRR} & \multicolumn{2}{l}{Novelty} & \multicolumn{2}{l}{Serendipity} & \multicolumn{2}{l}{Similarity} & \multicolumn{2}{l}{IntraList Diversity} \\
\cmidrule(lr){2-13}
                          & W          & r          & W          & r          & W            & r            & W              & r              & W              & r             & W             & r             \\
\midrule
MIND                      & 2E+08      & -0.98      & 5E+07      & -1         & 1E+10        & 0.35         & 8E+09          & 0.46           & 1E+10          & 0.08          & 3E+09         & -0.82         \\
K-Means Random            & 9E+06      & 0.39       & 1E+07      & 0.05       & 5E+09        & 0.67         & 4E+09          & 0.77           & 1E+10          & -0.3          & 8E+09         & -0.51         \\
K-Means Greedy            & 2E+06      & 0.79       & 7E+06      & 0.23       & 1E+10        & -0.24        & 1E+10          & -0.27          & 7E+09          & 0.56          & 7E+08         & -0.95         \\
History Avg.              & 7E+07      & -0.59      & 4E+07      & -0.8       & 1E+10        & -0.14        & 1E+10          & -0.07          & 8E+09          & 0.51          & 1E+09         & 0.93          \\
Uniform Random            & 2E+07      & 0.09       & 2E+07      & -0.19      & 1E+10        & 0.12         & 1E+10          & 0.24           & 1E+10          & 0.12          & 4E+09         & -0.73       \\ 
\bottomrule
\end{tabular}
}
\end{table}

% Please add the following required packages to your document preamble:
% \usepackage{multirow}
\begin{table}[h!]
\small
\centering
\caption{Wilcoxon signed-rank test results comparing \texttt{A2CG} with no diversity injection against baseline methods on normative diversity metrics. $W$ denotes the Wilcoxon statistic and $r$ the rank-biserial correlation effect size. All comparisons are statistically significant ($p < 0.0001$). Positive effect sizes indicate \texttt{A2CG} outperforms the baseline; negative values indicate the baseline outperforms \texttt{A2CG}. }
\begin{tabular}{lllllllllll}
\toprule
\multirow{2}{*}{Baseline Methods} & \multicolumn{2}{l}{Calibration} & \multicolumn{2}{l}{Activation} & \multicolumn{2}{l}{Representation} & \multicolumn{2}{l}{Alternative Voice} & \multicolumn{2}{l}{Fragmentation} \\
\cmidrule(lr){2-11}
                          & W               & r             & W              & r             & W                & r               & W                 & r                 & W                & r              \\
\midrule
MIND                      & 1E+10           & 0.21          & 8E+09          & 0.5           & 1E+10            & -0.21           & 1E+10             & -0.06             & 1E+10            & 0.29           \\
K-Means Random            & 6E+09           & 0.6           & 1E+08          & 0.99          & 1E+06            & 1               & 1E+10             & 0.23              & 0E+00            & 1              \\
K-Means Greedy            & 1E+10           & 0.29          & 3E+09          & 0.83          & 2E+09            & 0.89            & 2E+10             & -0.03             & 0E+00            & 1              \\
History Avg.              & 6E+09           & 0.61          & 1E+09          & -0.94         & 1E+10            & 0.28            & 7E+09             & 0.56              & 9E+09            & 0.44           \\
Uniform Random            & 1E+10           & 0.3           & 2E+09          & 0.86          & 3E+09            & 0.8             & 1E+10             & 0.16              & 4E+09            & 0.76          \\
\bottomrule
\end{tabular}
\end{table}

% Please add the following required packages to your document preamble:
% \usepackage{multirow}
\begin{table}[h!]
\small
\centering
\caption{Wilcoxon signed-rank test results comparing \texttt{A2CG} with no diversity injection against baseline methods on democratic models. $W$ denotes the Wilcoxon statistic and $r$ the rank-biserial correlation effect size. All comparisons are statistically significant ($p < 0.0001$). Positive effect sizes indicate \texttt{A2CG} outperforms the baseline; negative values indicate the baseline outperforms \texttt{A2CG}.}
\begin{tabular}{lllllllll}
\toprule
\multirow{2}{*}{Baseline Methods} & \multicolumn{2}{l}{Liberal} & \multicolumn{2}{l}{Participatory} & \multicolumn{2}{l}{Critical} & \multicolumn{2}{l}{Deliberative} \\
\cmidrule(lr){2-9}
                          & W            & r            & W               & r               & W             & r            & W               & r              \\
\midrule
MIND                      & 2E+10        & -0.02        & 8E+09           & 0.46            & 1E+10         & 0.21         & 9E+09           & -0.41          \\
K-Means Random            & 0E+00        & 1            & 1E+01           & -1              & 8E+06         & 1            & 0E+00           & -1             \\
K-Means Greedy            & 0E+00        & 1            & 8E+02           & -1              & 2E+09         & 0.87         & 0E+00           & -1             \\
History Avg.              & 1E+10        & -0.16        & 8E+09           & -0.44           & 8E+09         & -0.44        & 8E+09           & 0.49           \\
Uniform Random            & 1E+10        & 0.06         & 1E+10           & -0.19           & 2E+09         & 0.9          & 3E+09           & -0.82        \\
\bottomrule
\end{tabular}
\end{table}

% Please add the following required packages to your document preamble:
% \usepackage{multirow}
\begin{table}[h!]
\small
\centering
\caption{Wilcoxon signed-rank test results comparing \texttt{A2CG} variants against \texttt{A2CG} with no diversity injection (base) on traditional metrics. $W$ denotes the Wilcoxon statistic and $r$ the rank-biserial correlation effect size. All comparisons are statistically significant ($p < 0.0001$). Positive effect sizes indicate \texttt{A2CG} variants outperform the base; negative values indicate the base outperforms \texttt{A2CG} variants.}
\begin{tabular}{lllllllllllll}
\toprule
\multirow{2}{*}{\texttt{A2CG} Variants} & \multicolumn{2}{l}{AUC} & \multicolumn{2}{l}{MRR} & \multicolumn{2}{l}{Novelty} & \multicolumn{2}{l}{Serendipity} & \multicolumn{2}{l}{Similarity} & \multicolumn{2}{l}{Intra List Diversity} \\
\cmidrule(lr){2-13}
                               & W          & r          & W           & r         & W             & r           & W               & r             & W              & r             & W                   & r                  \\
\midrule
\rowcolor{lightgray}
\multicolumn{13}{c}{n=1} \\
10\%                           & 2E+06      & -0.52      & 1E+06       & 0.58      & 7E+09         & 0.18        & 1E+10           & 0.24          & 7E+09          & -0.44         & 5E+09               & 0.63               \\
25\%                           & 6E+06      & -0.21      & 4E+06       & 0.41      & 1E+10         & 0.24        & 1E+10           & 0.37          & 6E+09          & -0.63         & 3E+09               & 0.83               \\
50\%                           & 1E+07      & 0.05       & 9E+06       & 0.27      & 1E+10         & 0.32        & 8E+09           & 0.49          & 3E+09          & -0.78         & 1E+09               & 0.94               \\
75\%                           & 1E+07      & 0.11       & 1E+07       & 0.21      & 9E+09         & 0.38        & 6E+09           & 0.58          & 2E+09          & -0.85         & 4E+08               & 0.97               \\
100\%                          & 2E+07      & 0.12       & 2E+07       & 0.12      & 9E+09         & 0.43        & 6E+09           & 0.64          & 2E+09          & -0.88         & 2E+08               & 0.99               \\
\rowcolor{lightgray}
\multicolumn{13}{c}{n=2} \\
10\%                           & 3E+06      & -0.43      & 2E+06       & 0.52      & 7E+09         & 0.31        & 9E+09           & 0.41          & 6E+09          & -0.6          & 3E+09               & 0.79               \\
25\%                           & 9E+06      & -0.1       & 5E+06       & 0.41      & 8E+09         & 0.44        & 6E+09           & 0.58          & 3E+09          & -0.81         & 7E+08               & 0.95               \\
50\%                           & 1E+07      & 0.16       & 1E+07       & 0.34      & 7E+09         & 0.54        & 5E+09           & 0.71          & 2E+09          & -0.89         & 1E+08               & 0.99               \\
75\%                           & 2E+07      & 0.14       & 1E+07       & 0.3       & 6E+09         & 0.61        & 3E+09           & 0.77          & 1E+09          & -0.92         & 3E+07               & 1                  \\
100\%                          & 2E+07      & 0.17       & 2E+07       & 0.22      & 5E+09         & 0.66        & 3E+09           & 0.81          & 1E+09          & -0.93         & 1E+07               & 1                  \\
\rowcolor{lightgray}
\multicolumn{13}{c}{n=3} \\
10\%                           & 3E+06      & -0.41      & 2E+06       & 0.63      & 5E+09         & 0.44        & 6E+09           & 0.52          & 3E+09          & -0.75         & 1E+09               & 0.92               \\
25\%                           & 9E+06      & -0.07      & 5E+06       & 0.44      & 6E+09         & 0.55        & 5E+09           & 0.67          & 2E+09          & -0.87         & 2E+08               & 0.98               \\
50\%                           & 1E+07      & 0.16       & 1E+07       & 0.39      & 5E+09         & 0.65        & 3E+09           & 0.77          & 1E+09          & -0.92         & 3E+07               & 1                  \\
75\%                           & 2E+07      & 0.12       & 1E+07       & 0.37      & 4E+09         & 0.71        & 3E+09           & 0.82          & 1E+09          & -0.93         & 7E+06               & 1                  \\
100\%                          & 3E+07      & 0.13       & 2E+07       & 0.31      & 4E+09         & 0.75        & 2E+09           & 0.85          & 9E+08          & -0.94         & 3E+06               & 1                  \\
\rowcolor{lightgray}
\multicolumn{13}{c}{n=4} \\
10\%                           & 2E+06      & -0.59      & 2E+06       & 0.59      & 5E+09         & 0.47        & 6E+09           & 0.54          & 2E+09          & -0.81         & 6E+08               & 0.96               \\
25\%                           & 6E+06      & -0.3       & 5E+06       & 0.3       & 6E+09         & 0.58        & 5E+09           & 0.68          & 2E+09          & -0.9          & 1E+08               & 0.99               \\
50\%                           & 1E+07      & -0.09      & 9E+06       & 0.25      & 5E+09         & 0.66        & 4E+09           & 0.75          & 1E+09          & -0.93         & 1E+07               & 1                  \\
75\%                           & 1E+07      & -0.11      & 1E+07       & 0.26      & 4E+09         & 0.71        & 3E+09           & 0.79          & 1E+09          & -0.94         & 5E+06               & 1                  \\
100\%                          & 2E+07      & -0.11      & 2E+07       & 0.15      & 4E+09         & 0.74        & 3E+09           & 0.81          & 9E+08          & -0.94         & 3E+07               & 1                  \\
\rowcolor{lightgray}
\multicolumn{13}{c}{n=5} \\
10\%                           & 1E+06      & -0.71      & 1E+06       & 0.58      & 6E+09         & 0.42        & 7E+09           & 0.49          & 2E+09          & -0.86         & 3E+08               & 0.98               \\
25\%                           & 4E+06      & -0.44      & 5E+06       & 0.27      & 8E+09         & 0.48        & 7E+09           & 0.57          & 1E+09          & -0.91         & 5E+07               & 1                  \\
50\%                           & 7E+06      & -0.28      & 7E+06       & 0.2       & 7E+09         & 0.52        & 6E+09           & 0.61          & 1E+09          & -0.93         & 7E+06               & 1                  \\
75\%                           & 7E+06      & -0.34      & 9E+06       & 0.2       & 7E+09         & 0.56        & 6E+09           & 0.64          & 1E+09          & -0.94         & 3E+07               & 1                  \\
100\%                          & 7E+06      & -0.44      & 1E+07       & 0.02      & 6E+09         & 0.59        & 5E+09           & 0.66          & 9E+08          & -0.94         & 9E+09               & 0.45  \\
\bottomrule
\end{tabular}
\end{table}

% Please add the following required packages to your document preamble:
% \usepackage{multirow}
\begin{table}[h!]
\small
\centering
\caption{Wilcoxon signed-rank test results comparing \texttt{A2CG} variants against \texttt{A2CG} with no diversity injection (base) on normative diversity metrics. $W$ denotes the Wilcoxon statistic and $r$ the rank-biserial correlation effect size. All comparisons are statistically significant ($p < 0.0001$). Positive effect sizes indicate \texttt{A2CG} variants outperform the base; negative values indicate the base outperforms \texttt{A2CG} variants.}
\begin{tabular}{lllllllllll}
\toprule
\multirow{2}{*}{A2CG Variants} & \multicolumn{2}{l}{Calibration} & \multicolumn{2}{l}{Activation} & \multicolumn{2}{l}{Alternative Voice} & \multicolumn{2}{l}{Fragmentation} & \multicolumn{2}{l}{Representation} \\
\cmidrule(lr){2-11}
                               & W              & r              & W              & r             & W                 & r                 & W               & r               & W                & r               \\
\midrule
\rowcolor{lightgray}
\multicolumn{11}{c}{n=1} \\
10\%                           & 8E+09          & -0.14          & 5E+09          & -0.31         & 1E+10             & -0.1              & 1E+10           & -0.16           & 4E+09            & -0.19           \\
25\%                           & 1E+10          & -0.19          & 8E+09          & -0.4          & 1E+10             & -0.05             & 2E+10           & 0.01            & 9E+09            & -0.25           \\
50\%                           & 1E+10          & -0.24          & 8E+09          & -0.48         & 2E+10             & 0.02              & 1E+10           & -0.29           & 1E+10            & -0.33           \\
75\%                           & 1E+10          & -0.26          & 7E+09          & -0.53         & 1E+10             & 0.11              & 1E+10           & -0.37           & 9E+09            & -0.39           \\
100\%                          & 1E+10          & -0.28          & 7E+09          & -0.55         & 2E+10             & 0.01              & 8E+09           & -0.45           & 9E+09            & -0.43           \\
\rowcolor{lightgray}
\multicolumn{11}{c}{n=2} \\
10\%                           & 1E+10          & -0.19          & 6E+09          & -0.4          & 1E+10             & -0.14             & 1E+10           & -0.18           & 6E+09            & -0.31           \\
25\%                           & 1E+10          & -0.28          & 7E+09          & -0.52         & 1E+10             & -0.11             & 1E+10           & -0.2            & 9E+09            & -0.39           \\
50\%                           & 1E+10          & -0.33          & 6E+09          & -0.59         & 1E+10             & -0.05             & 9E+09           & -0.41           & 8E+09            & -0.51           \\
75\%                           & 1E+10          & -0.34          & 6E+09          & -0.62         & 1E+10             & 0.05              & 4E+09           & -0.74           & 6E+09            & -0.58           \\
100\%                          & 1E+10          & -0.34          & 6E+09          & -0.63         & 1E+10             & 0.05              & 2E+09           & -0.85           & 6E+09            & -0.63           \\
\rowcolor{lightgray}
\multicolumn{11}{c}{n=3} \\
10\%                           & 8E+09          & -0.27          & 5E+09          & -0.52         & 1E+10             & -0.18             & 1E+10           & -0.1            & 5E+09            & -0.46           \\
25\%                           & 1E+10          & -0.33          & 6E+09          & -0.6          & 1E+10             & -0.15             & 8E+09           & -0.48           & 6E+09            & -0.59           \\
50\%                           & 1E+10          & -0.36          & 5E+09          & -0.65         & 1E+10             & -0.08             & 5E+09           & -0.65           & 4E+09            & -0.73           \\
75\%                           & 1E+10          & -0.34          & 5E+09          & -0.66         & 1E+10             & 0.04              & 1E+09           & -0.91           & 3E+09            & -0.79           \\
100\%                          & 1E+10          & -0.29          & 5E+09          & -0.66         & 1E+10             & 0.21              & 3E+08           & -0.98           & 3E+09            & -0.83           \\
\rowcolor{lightgray}
\multicolumn{11}{c}{n=4} \\
10\%                           & 8E+09          & -0.32          & 5E+09          & -0.58         & 1E+10             & -0.2              & 2E+10           & 0.01            & 4E+09            & -0.63           \\
25\%                           & 1E+10          & -0.34          & 6E+09          & -0.64         & 1E+10             & -0.15             & 1E+10           & -0.39           & 4E+09            & -0.75           \\
50\%                           & 1E+10          & -0.32          & 5E+09          & -0.66         & 1E+10             & -0.04             & 2E+09           & -0.85           & 3E+09            & -0.83           \\
75\%                           & 1E+10          & -0.24          & 6E+09          & -0.64         & 1E+10             & 0.14              & 3E+08           & -0.98           & 2E+09            & -0.85           \\
100\%                          & 1E+10          & -0.12          & 7E+09          & -0.57         & 7E+09             & 0.53              & 8E+06           & -1              & 2E+09            & -0.84           \\
\rowcolor{lightgray}
\multicolumn{11}{c}{n=5} \\
10\%                           & 9E+09          & -0.31          & 5E+09          & -0.6          & 1E+10             & -0.17             & 1E+10           & -0.04           & 4E+09            & -0.7            \\
25\%                           & 1E+10          & -0.29          & 6E+09          & -0.6          & 1E+10             & -0.07             & 8E+09           & -0.49           & 4E+09            & -0.73           \\
50\%                           & 1E+10          & -0.19          & 7E+09          & -0.53         & 1E+10             & 0.11              & 2E+09           & -0.87           & 5E+09            & -0.69           \\
75\%                           & 2E+10          & -0.02          & 9E+09          & -0.39         & 1E+10             & 0.35              & 3E+08           & -0.98           & 7E+09            & -0.57           \\
100\%                          & 1E+10          & 0.21           & 2E+10          & 0.03          & 3E+09             & 0.8               & 5E+06           & -1              & 1E+10            & -0.34          
\end{tabular}
\end{table}

% Please add the following required packages to your document preamble:
% \usepackage{multirow}
\begin{table}[h!]
\small
\centering
\caption{Wilcoxon signed-rank test results comparing \texttt{A2CG} variants against \texttt{A2CG} with no diversity injection (base) on democratic models. $W$ denotes the Wilcoxon statistic and $r$ the rank-biserial correlation effect size. All comparisons but one (marked with *) are statistically significant ($p < 0.0001$). Positive effect sizes indicate \texttt{A2CG} variants outperform the base; negative values indicate the base outperforms \texttt{A2CG} variants.}
\begin{tabular}{lllllllll}
\toprule
\multirow{2}{*}{A2CG Variants} & \multicolumn{2}{l}{Liberal} & \multicolumn{2}{l}{Participatory} & \multicolumn{2}{l}{Critical} & \multicolumn{2}{l}{Deliberative} \\
\cmidrule(lr){2-9}
                               & W          & r              & W                & r              & W             & r            & W               & r              \\
\midrule
\rowcolor{lightgray}
\multicolumn{9}{c}{n=1} \\
10\%                           & 1E+10      & -0.08          & 1E+10            & 0.17           & 1E+10         & -0.27        & 1E+10           & 0.26           \\
25\%                           & 1E+10      & 0.13           & 1E+10            & 0.11           & 1E+10         & -0.36        & 1E+10           & 0.24           \\
50\%                           & 2E+10      & -0.01          & 1E+10            & 0.21           & 9E+09         & -0.43        & 9E+09           & 0.4            \\
75\%                           & \textbf{2E+10}*      & \textbf{0}*     & 1E+10            & 0.21           & 8E+09         & -0.45        & 9E+09           & 0.4            \\
100\%                          & 2E+10      & -0.02          & 1E+10            & 0.23           & 7E+09         & -0.55        & 8E+09           & 0.51           \\
\rowcolor{lightgray}
\multicolumn{9}{c}{n=2} \\
10\%                           & 1E+10      & -0.04          & 1E+10            & 0.22           & 8E+09         & -0.42        & 1E+10           & 0.37           \\
25\%                           & 1E+10      & 0.07           & 1E+10            & 0.21           & 7E+09         & -0.54        & 8E+09           & 0.47           \\
50\%                           & 1E+10      & 0.05           & 1E+10            & 0.25           & 6E+09         & -0.63        & 7E+09           & 0.56           \\
75\%                           & 1E+10      & -0.15          & 1E+10            & 0.32           & 5E+09         & -0.65        & 6E+09           & 0.63           \\
100\%                          & 1E+10      & -0.22          & 1E+10            & 0.35           & 5E+09         & -0.69        & 5E+09           & 0.67           \\
\rowcolor{lightgray}
\multicolumn{9}{c}{n=3} \\
10\%                           & 1E+10      & 0.08           & 1E+10            & 0.24           & 6E+09         & -0.58        & 9E+09           & 0.42           \\
25\%                           & 1E+10      & -0.06          & 9E+09            & 0.4            & 4E+09         & -0.71        & 5E+09           & 0.65           \\
50\%                           & 1E+10      & -0.06          & 9E+09            & 0.44           & 3E+09         & -0.79        & 5E+09           & 0.69           \\
75\%                           & 1E+10      & -0.29          & 7E+09            & 0.52           & 3E+09         & -0.79        & 4E+09           & 0.72           \\
100\%                          & 8E+09      & -0.47          & 6E+09            & 0.59           & 4E+09         & -0.75        & 5E+09           & 0.68           \\
\rowcolor{lightgray}
\multicolumn{9}{c}{n=4} \\
10\%                           & 1E+10      & 0.21           & 1E+10            & 0.31           & 4E+09         & -0.72        & 9E+09           & 0.44           \\
25\%                           & 1E+10      & 0.05           & 8E+09            & 0.48           & 3E+09         & -0.81        & 5E+09           & 0.65           \\
50\%                           & 1E+10      & -0.25          & 7E+09            & 0.57           & 2E+09         & -0.84        & 4E+09           & 0.75           \\
75\%                           & 7E+09      & -0.52          & 5E+09            & 0.64           & 3E+09         & -0.8         & 4E+09           & 0.74           \\
100\%                          & 4E+09      & -0.75          & 5E+09            & 0.68           & 7E+09         & -0.56        & 7E+09           & 0.53           \\
\rowcolor{lightgray}
\multicolumn{9}{c}{n=5} \\
10\%                           & 1E+10      & 0.18           & 1E+10            & 0.38           & 3E+09         & -0.77        & 8E+09           & 0.49           \\
25\%                           & 1E+10      & -0.06          & 8E+09            & 0.47           & 3E+09         & -0.78        & 5E+09           & 0.65           \\
50\%                           & 1E+10      & -0.38          & 8E+09            & 0.49           & 5E+09         & -0.67        & 5E+09           & 0.65           \\
75\%                           & 5E+09      & -0.67          & 7E+09            & 0.54           & 9E+09         & -0.4         & 7E+09           & 0.55           \\
100\%                          & 1E+09      & -0.93          & 5E+09            & 0.65           & 7E+09         & 0.52         & 2E+10           & 0.02          \\
\bottomrule
\end{tabular}
\end{table}

\FloatBarrier
\section{User-Content Archetype}
\label{app:archetype}
\begin{figure}[h!]
    \centering
    \includegraphics[width=1\linewidth]{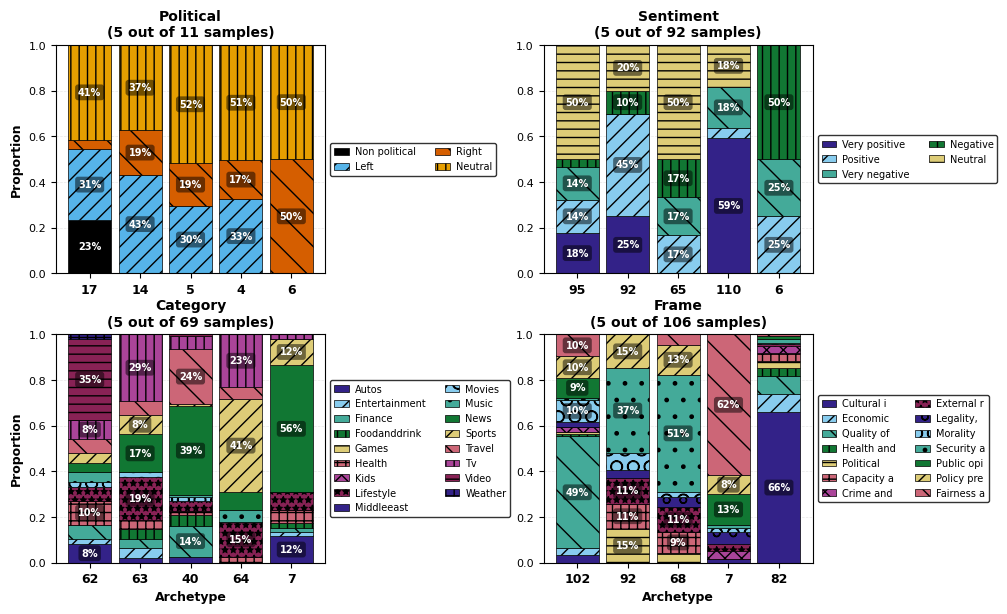}
    \caption{User-Content Archetypes for four aspects (Political, Sentiment, Category and Frame)}
    \label{fig:placeholder}
\end{figure}

\FloatBarrier
\end{document}